\documentclass[
 aip,
 jmp,%
 amsmath,amssymb,
 reprint,%
 raggedbottom,%
]{revtex4-2}

\usepackage{graphicx}% Include figure files
\usepackage{dcolumn}% Align table columns on decimal point
\usepackage{bm}% bold math
\usepackage{soul}
\usepackage{color}
\renewcommand{\d}{\mathrm{d}}
\usepackage{printlen}
\uselengthunit{cm}

\begin{document}

\preprint{AIP/123-QED}

\title[Girsanov Reweighting with CP2K
]{Implementation of Girsanov Reweighting in CP2K}% Force line breaks with \\

\author{Sascha Jähnigen (ORCID iD: \href{https://orcid.org/0000-0002-2480-3313}{0000-0002-2480-3313})}
\affiliation{ 
Freie Universität Berlin, Department of Biology, Chemistry and Pharmacy, Arnimallee 22, 14195 Berlin
}%
\author{Bettina G. Keller (ORCID iD: \href{https://orcid.org/0000-0002-7051-0888}{0000-0002-7051-0888})}%
\email{bettina.keller@fu-berlin.de}
\affiliation{ 
Freie Universität Berlin, Department of Biology, Chemistry and Pharmacy, Arnimallee 22, 14195 Berlin
}%
\date{\today}% It is always \today, today,

\begin{abstract}
Dynamical reweighting of path measures is a powerful approach for accurately evaluating slow molecular processes using modified potential energy surfaces used in enhanced sampling methods.
Integrating this reweighting framework into the CP2K electronic-structure and molecular-dynamics (MD) software package delivers a robust and widely applicable tool for metadynamics, uncertainty quantification, and force-field optimisation based on both \textit{ab initio} and classical MD simulations.
Based on the Girsanov theorem for stochastic dynamical systems, the method is adapted to the Bussi–Donadio–Parrinello velocity-rescaling scheme. This scheme is accessible through the CSVR thermostat and can be interpreted as a Langevin OVRVO/OBABO update scheme requiring two random numbers per integration step.
Comprehensive implementation details are provided, including a complete overview of the modified CP2K modules involved in the MD cycle and the computation of reweighting factors.
The framework supports multiple sources of bias potentials, enabling reweighting via the PLUMED interface or CP2K-native external potentials and restraints.
The feasibility of the implementation is demonstrated through rerun benchmarks, dynamical reweighting of Markov state models (MSMs) and computation of transport properties based on CP2K trajectories, showing excellent agreement with reference simulations.
\end{abstract}

%Use showkeys class option if keyword

\maketitle

% \section*{\hl{Notes}}
% \begin{itemize}
%     \item Visualise calculation of TCF 
%     \item Figures: show the file/commnunication CP2K network and where the implementation goes
% \end{itemize}

\section{\label{sec:Introduction}Introduction}

Many molecular processes are dominated by rare events, where the waiting time for the event to occur is orders of magnitude longer than the transition time itself.
Examples include chemical reactions, conformational rearrangements in proteins, ligand binding and unbinding, nucleation events in self-assembly, or ion permeation through membrane channels.
However, complex molecular systems do not typically exhibit clean timescale separation. 
Instead, they display metastability: the system dwells in one of several metastable conformations for extended periods.
The exit time from these states is much longer than the rapid transition times between neighbouring metastable states. 
Moreover, each metastable state may itself contain substates that interconvert on timescales that are not well separated from the exit dynamics.
From a modelling perspective, the difficulty is that the long waiting times of rare events exceed the accessible molecular-dynamics (MD) simulation timescales, making direct sampling of the full process infeasible.\cite{bolhuis_transition_2002}
Enhanced sampling methods address this challenge by generating trajectories at modified thermodynamic conditions for example by applying bias potentials or increasing the temperature.\cite{B027,G185,G195}
It is well established that reweighting such simulations can be used to recover stationary distributions and ensemble averages in the target ensemble.\cite{chipot2007free, stoltz2010free}
More recently, dynamical reweighting methods\cite{keller2024dynamical} have been developed to reconstruct kinetic properties such as time-correlation functions (TCFs) and state-to-state transition probabilities of the target ensemble.
Most existing approaches rely on an assumed effective dynamical model rather than reweighting the sampled trajectories themselves, which limits their accuracy and can introduce systematic bias.
In contrast, Girsanov reweighting operates directly on full-dimensional MD paths at the resolution of the integrator, making it a particularly accurate and model-free route to dynamical reweighting.\cite{donati2017girsanov}
The Girsanov theorem\cite{G182} is a fundamental result of stochastic analysis: it specifies the conditions under which a trajectory generated under a simulation ensemble has a finite probability under a target ensemble.
If these conditions are met, one can compute the ratio of path probabilities between the two ensembles, i.e.~the relative path weight, and use it to reweight path ensemble averages, thereby recovering kinetic properties of the target dynamics. 
Although Girsanov reweighting has been explored in the context of molecular simulations \cite{mazonka1998computing, zuckerman1999dynamic, athenes2004path, xing2006calculation, adib2008stochastic}, several practical obstacles have prevented its routine use.
First, the relative path weight decays exponentially with path length, restricting reweighting to trajectories far shorter than the relevant molecular timescales. 
Second, the path must be recorded at the full time resolution of the MD integrator, which is prohibitive for high-dimensional systems.
Third, earlier formulations were derived for overdamped Langevin dynamics, whereas realistic MD simulations follow underdamped dynamics, making direct application inconsistent \cite{kieninger2021path}.
To overcome the first obstacle, long-timescale behaviour must be inferred from short trajectories. 
This can be achieved by parameterising the transfer operator $\mathcal{T}_{\tau}$, which propagates probability densities over a short time-interval $\tau$. 
Markov state models (MSMs) \cite{prinz2011markov, schutte1999direct, lemke2016density} provide a finite-dimensional approximation of this transfer operator by partitioning configuration space and estimating the resulting transition matrix from trajectories of length $\tau$, where $\tau$ is much shorter than the slow molecular timescales.
This makes Girsanov reweighting feasible at the level of MSM transition probabilities \cite{schutte2015markov, donati2017girsanov, R086}.
An alternative strategy avoids reweighting full transition probabilities and instead reweights only the flux across transition regions, which can be estimated from short transition paths \cite{bolhuis2023optimizing}.
The second obstacle can be overcome by computing the Girsanov reweighting factor on-the-fly during the simulation and only storing the accumulated weight at the times when the trajectory is written to disk \cite{donati2017girsanov}.
This is feasible if the target potential is known at the beginning of the simulation, which is naturally the case when reweighting a biased simulation to the underlying unbiased molecular potential \cite{donati2018girsanov}. 
Moreover, using the MD integrator equations, the relative path weight can be expressed in terms of the random number sequence that was used to generate the simulated trajectory, and the hypothetical sequence that would produce the same path under the target dynamics.
This eliminates the need to record the full trajectory at integrator resolution \cite{kieninger2021path}.
Finally, underdamped Langevin dynamics are commonly realised in MD simulation packages through Langevin splitting schemes \cite{G088, G213, leimkuhler2013robust, fass2018quantifying, zhang2019unified, kieninger2022gromacs}, often termed stochastic thermostats.
A subset of these schemes satisfies the assumptions required for Girsanov reweighting, allowing explicit formulas for path-weight factors\cite{X001}.
The practical implementation of these reweighting factors in the MD engine OpenMM is presented in Ref.~\citenum{M076}.

Recent applications of Girsanov reweighting include kinetic force-field optimisation \cite{bolhuis2023optimizing}, catalytic cycle acceleration \cite{bolhuis2025optimal}, and machine-learning of slow collective variables \cite{shmilovich2023girsanov}.
These studies, however, rely on empirical force fields or model potentials, which cannot describe chemical reactions involving the breaking and formation of covalent bonds.
Ab initio molecular dynamics (AIMD) based on density functional theory (DFT) provides such a description, but the associated computational cost drastically limits accessible simulation times.\cite{B006} 
As a result, the timescale problem is even more pronounced than in simulations with empirical force fields, while chemical reactions typically occur on slower timescales. 
Consequently, simulations of chemical reactions almost invariably rely on enhanced sampling methods. 

Once an unbiasing scheme of dynamic properties has been established, 
the scope of reweighting can be extended beyond MSMs to other dynamic properties that are defined on a trajectory and not individual states.
The most notable example of this is transport properties such as diffusion. 
Also here, a biasing force can enhance the transport of particles (or heat, etc.), the quantification of which would otherwise require long simulation times, particularly if the process is linked to a rare event.

The aim of the present contribution is therefore to make Girsanov dynamical reweighting available at the AIMD level by implementing the required path-weight factors into the open source package CP2K\cite{vandevondele2005quickstep, kuhne2020cp2k,CP2K,iannuzzi_cp2k_2025}.
CP2K offers a wide range of computational methods and is optimised for large-scale applications on high-performance computing systems.
In addition to quantum and DFT methods, it supports classical force fields and machine-learning interatomic potentials (MLIPs), and it allows these approaches to be combined in hybrid schemes such as QM/MM.
With this work, we integrate the established framework of Girsanov reweighting with the existing capabilities of CP2K, and describe in detail how this implementation can be accessed and applied to perform dynamical reweighting of MSMs and transport properties, both of which rely on TCFs.

% ----------------------------------------------------------------------------
\section{Theory}
% ----------------------------------------------------------------------------
\label{sec:Theory}

% ----------------------------------------------------------------------------
\subsection{The Canonical Ensemble and Underdamped Langevin Dynamics}
Consider a system with $N$ atoms with positions $\bm{q} \in \mathbb{R}^{3N}$ and velocities $\dot{\bm{q}} \in \mathbb{R}^{3N}$. 
The Hamiltonian of the system subject to a potential $V(\bm{q})$ is
%\begin{equation}
$
H(\bm{q},\dot{\bm{q}}) = K(\dot{\bm{q}}) + V(\bm{q}),
$
%\end{equation}
with the kinetic energy
$
K(\dot{\bm{q}})= \sum_{i=1}^{3N} \frac{1}{2} m_i \,\dot{q}_{i}^2, 
\label{eq:E_kin}
$
where $m_i$ is the particle mass associated to the $i$th degree of freedom.

A closed system that is in thermal contact with a heat bath at temperature $T$ is described by the canonical ensemble and can be evolved according to underdamped Langevin dynamics,
\begin{align}
\d{}q_{i} &= \dot{q}_{i} \, \d{}t, \cr
\d{}\dot{q}_{i} &= \left(-\frac{\nabla_{i} V(\bm{q})}{m_i}  - \xi \dot{q}_{i} \right)\, \d{}t + \sqrt{2 \xi \frac{k_B T}{m_i} } \; \d{}W_i,
\label{eq:ULE}
\end{align}
where $\xi > 0$ is the collision or friction rate (in units of time$^{-1}$) and $k_B$ is the Boltzmann constant.
$W_i(t)$ are independent standard Wiener processes, one for each degree of freedom $i$ of the system.
% , indexed by the flattened global index $j\equiv(i,\alpha)$ (using row-major ordering). \hl{index of Wiener process normally indicates the time (step), here particle = problem?}
The increments $\d{}W_i$ are Gaussian random variables with zero mean and variance $dt$, i.e.
$\d{}W_i(t) \sim \mathcal{N}(0,dt)$.
They satisfy the average $\langle \d{}W_i(t)\rangle = 0$
and the correlation $\langle \d{}W_i(t)\,\d{}W_{i'}(t')\rangle = \delta_{ii'}\delta(t-t')\d{}t$, where
$\delta_{ii'}$ and $\delta(t-t')$ denote the Kronecker delta and the Dirac delta function, respectively.

Hence, the force acting on the $i$th degree of freedom, $F_i\d{} t = m_i\mathrm{d}\dot{q}_i$ (lower line in Eq.~\ref{eq:ULE}) is composed of deterministic terms depending on $\d{}t$ -- the conservative force from the potential gradient (first term) and the non-conservative friction force (second term) -- and stochastic terms depending on $\d{}W_i$.
The energy dissipation due to friction and the energy input from stochastic forces are balanced on average according to the fluctuation--dissipation theorem~\cite{kubo1966fluctuation}, thereby maintaining the canonical distribution and a constant equilibrium temperature.

% ----------------------------------------------------------------------------
\subsection{Integrating the Equations of Motion}
Numerical integration of Eq.~\ref{eq:ULE} yields a time-discretised trajectory of successive states $\bm{x}_0, \bm{x}_1, \dots$  separated by a fixed time step $\Delta t$.
Each state vector $\bm{x}_k = (\bm{q}_k,\dot{\bm{q}}_k) \in \Gamma \subset \mathbb{R}^{6N}$ represents the system configuration and velocities at time $t = k\cdot\Delta t$, where $\Gamma$ denotes the phase space.

Langevin integrators can be formulated using the operator-splitting method, in which Eq.~\ref{eq:ULE} is expressed as a sum of distinct vector fields \cite{matthews2015molecular, leimkuhler2013robust, matthews2015molecular},
\begin{equation}
    \left(
    \begin{array}{c}
        \d{q}_i\\ \d\dot{q}_i
    \end{array}
    \right)
    = 
    \underbrace{
    \left(\begin{array}{c}
        \dot{q}_i \d t \\ 0 
    \end{array} \right) }_{R}
    +
    \underbrace{
    \left(\begin{array}{c}
        0 \\ - \frac{\nabla_i V(\bm{q})}{m_i} \d t
    \end{array} \right)}_{V} 
    +
    \underbrace{     
    \left(\begin{array}{c}
        0 \\ -\xi \dot{q}_i \d t + \sqrt{2\xi \frac{k_B T}{m_i}} \, \d{W}_i
    \end{array} \right)}_{O}.
\label{eq:ULE_split}         
\end{equation}
The components $R$, $V$ and $O$ correspond, respectively, to the position update, the conservative velocity update, and the velocity update due to friction and random forces (Ornstein–Uhlenbeck process).
Individual integration of the three components gives rise to the corresponding update operators,\cite{matthews2015molecular}
\begin{subequations}
\begin{align}
\mathcal{R}\begin{pmatrix} q_{i,k} \\ \dot{q}_{i,k} \end{pmatrix} 
    % &= \begin{pmatrix} q_{i,k} + \Delta t \,\dot{q}_{i,k} \\ \dot{q}_{i,k} \end{pmatrix} 
    &= \begin{pmatrix} q_{i,k} + a\,\dot{q}_{i,k} \\ \dot{q}_{i,k} \end{pmatrix}, \label{eq:update_operators_R}\\
\mathcal{V}\begin{pmatrix} q_{i,k} \\ \dot{q}_{i,k} \end{pmatrix} 
    % &= \begin{pmatrix} q_{i,k} \\ \dot{q}_{i,k} -\frac{\nabla_i V(\bm{q}_k)}{m_i} \Delta t \end{pmatrix} 
     &= \begin{pmatrix} q_{i,k} \\ \dot{q}_{i,k} + b_i(\bm{q}_k) \end{pmatrix}, \\
\mathcal{O}\begin{pmatrix} q_{i,k} \\ \dot{q}_{i,k} \end{pmatrix} 
    % &= \begin{pmatrix} q_{i,k} \\ \exp\left({-\xi\Delta t}\right)\, \dot{q}_{i,k} + \sqrt{\frac{k_BT}{m_i}(1-\exp\left({-2\xi\Delta t}\right))} \,\eta_{i,k} \end{pmatrix} 
    &= \begin{pmatrix} q_{i,k} \\ d \,\dot{q}_{i,k} + f\, \eta_{i,k} \end{pmatrix},
\end{align}%
\label{eq:update_operators}%
\end{subequations}%
for which the following parameters have been defined:
\begin{align}
    a       &= \Delta t,\cr
    b_i(\bm{q})  &= -\frac{\nabla_i V(\bm{q})}{m_i}\Delta t,\cr
    d       &= \exp\left({-\xi\Delta t}\right),\cr
    f       &= \sqrt{\frac{k_BT}{m_i}(1-\exp\left({-2\xi\Delta t}\right)).} 
\label{eq:parameters}
\end{align}
The Ornstein–Uhlenbeck update ($\mathcal{O}$) introduces independent Gaussian random numbers 
$\eta_{i,k} \sim \mathcal{N}(0, 1)$ drawn at each time step to sample the Wiener process.
The updated state $\bm{x}_{k+1}$ is obtained by successively applying the operators in Eq.~\ref{eq:update_operators} to $\bm{x}_k$, and the various integration schemes differ only in the order in which these operators are applied.\cite{X001}
According to the Strang splitting principle, symmetric operator sequences, where certain operators are applied twice for half a timestep, provide a particularly accurate approximation of the underlying continuous Langevin dynamics.
The symmetric Langevin splitting scheme $\mathcal{O}'\mathcal{V}'\mathcal{R}\mathcal{V}'\mathcal{O}'$ is widely used and is also implemented in CP2K through the CSVR thermostat (see Section~\ref{sec:implementation}).\cite{G088,iannuzzi_cp2k_2025} The prime superscript indicates that the corresponding update operator is applied for half a time step, i.e., using $\Delta t/2$ instead of $\Delta t$ in Eq.~\ref{eq:parameters}.

%

% ----------------------------------------------------------------------------
\subsection{Path Reweighting}
\label{sec:pathReweighting}

Let $\mathbf{x} = \{\bm{x}_0, \bm{x}_1, \bm{x}_2 \dots, \bm{x}_n\} \in \mathcal{S}$ be a time-discretised path of length $n \cdot \Delta{}t$, where $n$ is the number of time steps.
$\mathcal{S} = \Gamma^{n+1}$ is the path space that contains all possible paths of length $n\cdot\Delta t$.
In the $\mathcal{O}'\mathcal{V}'\mathcal{R}\mathcal{V}'\mathcal{O}'$ Langevin splitting scheme, two independent Gaussian random numbers are drawn per timestep (one in each $\mathcal{O}$ half-step), which makes the update from $\bm{x}_k$ to $\bm{x}_{k+1}$ stochastic, with a transition probability density $p(\bm{x}_{k+1}\,|\,\bm{x}_k)$.
The path probability density, $\mathcal{P}[\mathbf{x}]$, for observing a specific path $\mathbf{x}$ can be expressed as the product of one-step transition probabilities
\begin{equation}
    \mathcal{P}[\mathbf{x}] = p(\bm{x}_0) \cdot\mathcal{P}[\bm{x}_1\dots{}\bm{x}_n|\bm{x}_0] = p(\bm{x}_0) \cdot\prod_{k=0}^{n-1} p(\bm{x}_{k+1}|\bm{x}_k),
    \label{eq:chapman–kolmogorov}
\end{equation}
with the probability density of the initial state $p(\bm{x}_0)$, defined through the Maxwell-Boltzmann-distribution,
\begin{equation}
p(\bm{x}_0) \equiv p(\bm{q}_0,\dot{\bm{q}}_0)
= \frac{1}{Z} 
\exp\left( -\frac{V(\bm{q_0})}{k_BT} \right)
\prod_{i=1}^{3N} 
\frac{1}{\sqrt{2\pi k_B T m_i}}
\exp\left( -\frac{m_i \dot{q}_{i,0}^2 }{2 k_B T} \right),
    \label{eq:Maxwell_Boltzmann}
\end{equation}
with the classical partition function $Z$.
Since the deterministic sub-steps $\mathcal{V}$ and $\mathcal{R}$ are determined by the current state $\bm{x}_k$, the update to state $\bm{x}_{k+1}$ is fully specified by the two Gaussian random number vectors $\bm{\eta}_k^{(1)}\in \mathbb{R}^{3N}$ and $\bm{\eta}_k^{(2)}\in \mathbb{R}^{3N}$, drawn in the two $\mathcal{O}$ half-steps. 
Accordingly, the transition probability density $p(\bm{x}_{k+1}|\bm{x}_k)$ is equal to the probability density of drawing this pair of independent Gaussian random numbers, \cite{kieninger2021path}

\begin{equation}
    p(\bm{x}_{k+1}|\bm{x}_k) 
    = p\!\left(\bm{\eta}_k^{(1)}\right) p\!\left(\bm{\eta}_k^{(2)}\right) 
    = \prod_{i=1}^{3N} \frac{1}{2\pi}\exp\left(- \frac{{\eta_{i,k}^{(1)}}^2}{2}\right)
    \cdot \exp\left(- \frac{{\eta_{i,k}^{(2)}}^2}{2}\right)\, .
    \label{eq:white_noise}
\end{equation}
The path probability density is normalised, which defines a path integral measure $\int_{\mathcal{S}} \mathcal{D}\mathbf{x}$, such that 
\begin{equation}
    \int_{\mathcal{S}} \mathcal{D}\mathbf{x} \, \mathcal{P}[\mathbf{x}] = 
    \int_\Gamma\d{}\bm{x}_0 \int_\Gamma \d{}\bm{x}_1 \cdots \int_\Gamma \d{}\bm{x}_n
    \,\mathcal{P}[\mathbf{x}]  = 1.
    \label{eq:path_measure}
\end{equation}
where $\Gamma$ represents the configuration space at each time step, 
and $\mathcal{D}\mathbf{x} = \prod_{k=0}^n \mathrm{d}\bm{x}_k$.

Let $s:\mathcal{S} \rightarrow\mathbb{R}$ be a path observable, i.e. a function that assigns a real-valued number to each time-discretised path.
The expected value of $s$ over the path ensemble is then
\begin{align}
    \langle s \rangle &= \int_{\mathcal{S}} \mathcal{D}\mathbf{x} \, \mathcal{P}[\mathbf{x}]\, s[\mathbf{x}]
    = \lim_{|A| \rightarrow \infty}  \frac{1}{|A|}\, \sum_{\mathbf{x} \in A} s[\mathbf{x}].
    \label{eq:expected_path}
\end{align}
The second equality of Eq.~\ref{eq:expected_path} defines an estimator of the path expected value based on the arithmetic mean over a set of paths {$A \subset \mathcal{S}$} extracted from a numerical simulation under the assumption of ergodicity.
The simulations are carried out at the simulation potential $V(\bm{q})$. 
Now assume a target potential energy function,
\begin{equation}
    \widetilde{V}(\bm{q}) = {V}(\bm{q}) + U(\bm{q}),
    \label{eq:V_tilde}
\end{equation}
which deviates from $V(\bm{q})$ by a  perturbation potential $U(\bm{q})$.
In the following, the tilde symbol ($\,\widetilde{\;}\,$) is used to indicate that a quantity has been defined with respect to the target potential.
The objective is to reweight the estimator in Eq.~\ref{eq:expected_path} such that the path-ensemble average $\widetilde{\langle s \rangle}$ corresponding to the target potential $\widetilde{V}(\bm{q})$ is obtained from the path sample obtained at the simulation potential $V(\bm{q})$.
A necessary condition for reweighting the estimator is that the sampled path space obtained under the potential ${V}(\bm{q})$, includes all paths that are theoretically accessible under the modified potential $\widetilde{V}(\bm{q})$ (absolute continuity).
Hence, the probability measures based on either $\widetilde{\mathcal{P}}[\mathbf{x}]$ or $\mathcal{P}[\mathbf{x}]$ can be defined on $\mathcal{S}$ through Eq.~\ref{eq:path_measure}, because $\text{supp}(\widetilde{\mathcal{P}}) \subseteq \text{supp}(\mathcal{P})$.
In this case, the Radon–Nikodym theorem ensures the existence of a reweighting factor:
\begin{equation}
    \frac{\widetilde{\mathcal{P}}[\mathbf{x}]}{\mathcal{P}[\mathbf{x}]} 
    = \frac{\widetilde{p}(\bm{x}_0)} {p(\bm{x}_0)} \cdot \frac{\widetilde{\mathcal{P}}[\bm{x}_1\dots{}\bm{x}_n|\bm{x}_0]}{\mathcal{P}[\bm{x}_1\dots{}\bm{x}_n|\bm{x}_0]}
    = g(\bm{x}_0) \cdot M[\mathbf{x}|\bm{x}_0], 
    \label{eq:Radon-Nikodym}
\end{equation}
where $M[\mathbf{x}|\bm{x}_0]$ denotes the ratio of conditional path probability densities,
and $g(\bm{x}_0)$ represents the static reweighting factor associated with the initial state.
The path expected value of $s$ at the target potential $\widetilde{V}$ reads,\cite{donati2017girsanov}
\begin{align}
    \widetilde{\langle s \rangle} &= \int_{\mathcal{S}} \mathcal{D}\mathbf{x} \, \widetilde{\mathcal{P}}[\mathbf{x}]\, s[\mathbf{x}]
    = \int_{\mathcal{S}} \mathcal{D}\mathbf{x} \; g(\bm{x}_0) M[\mathbf{x}|\bm{x}_0]
    \,\mathcal{P}[\mathbf{x}]\, s[\mathbf{x}] \cr
    &= \lim_{|A| \rightarrow \infty}  \frac{1}{|A|}\, \sum_{\mathbf{x} \in A}
    \,g(\bm{x}_0)\, M[\mathbf{x}|\bm{x}_0] \,s[\mathbf{x}],
    \label{eq:reweighted_expected_path}
\end{align}
where in the last line the estimator is defined according to Eq.~\ref{eq:expected_path}.

\begin{figure}
    \includegraphics[width=\textwidth]{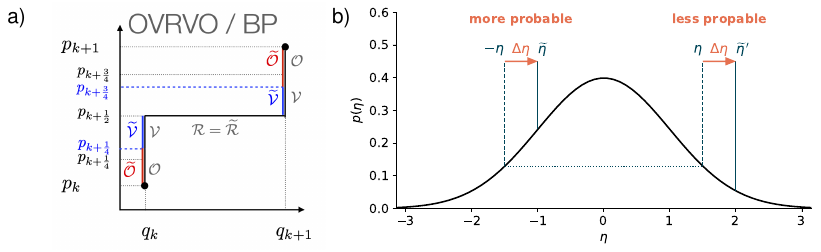}
    \caption{
    a) Overlay of congruent update graphs from $(q_{i,k}, p_{i,k})$ to $(q_{i,k+1}, p_{i,k+1})$ for the $\mathcal{O'V'RV'O'}$ integrator, comparing the simulation potential $V(\bm{q}_k)$ with the target potential $\widetilde{V}(\bm{q}_k)$;\cite{X001,M076}
    b) Impact of the location of the support point on $\mathcal{N}(0,1)$ for the calculation of $\widetilde{\eta}$: Depending on the sign of $\eta$, a one-step transition can become more or less probable upon reweighting.} 
    \label{fig:OBABO}
\end{figure} 

% ----------------------------------------------------------------------------
\subsection{Girsanov Reweighting Factors for $\mathcal{O}'\mathcal{V}'\mathcal{R}\mathcal{V}'\mathcal{O}'$}
\label{sec:GirsanovReweightingFactor}
The $\mathcal{O}'\mathcal{V}'\mathcal{R}\mathcal{V}'\mathcal{O}'$ scheme fulfils absolute continuity, and thus one can  derive concrete expressions for the reweighting factors in the Radon-Nikodym derivative (Eq.~\ref{eq:Radon-Nikodym}). \cite{X001}
The static reweighting factor is obtained by substituting Eqs.~\ref{eq:Maxwell_Boltzmann} into Eq.~\ref{eq:Radon-Nikodym}
\begin{equation}
    g(\bm{x}_0) = \frac{\widetilde{p}(\bm{x}_0)} {p(\bm{x}_0)} = \frac{Z}{\widetilde{Z}}\exp\left(-\frac{U(\bm{q}_0)}{k_BT} \right).
    \label{eq:g}
\end{equation}
Using Eq.~\ref{eq:V_tilde}, $g(\mathbf{x}_0)$ simplifies to a function that only depends on the perturbation potential $U(\bm{q}_0)$. 
The partition function ratio $\frac{Z}{\widetilde{Z}}$ cancels in the final estimator\cite{donati2017girsanov} and does not need to be calculated (see Section~\ref{sec:observables}).

The relative path-weights for the conditional paths are obtained by substituting Eqs.~\ref{eq:chapman–kolmogorov} and \ref{eq:white_noise} into Eq.~\ref{eq:Radon-Nikodym}
\begin{equation}
    M[\mathbf{x} | \bm{x}_0 ]
    =  \prod_{k=0}^{n-1} \prod_{i=1}^{3N}\exp\left(
    - \eta^{(1)}_{i,k}\cdot\Delta\eta^{(1)}_{i,k} 
    - \frac 1 2 \Delta{\eta^{(1)}_{i,k}}^2 
    \;-\; \eta^{(2)}_{i,k}\cdot\Delta\eta^{(2)}_{i,k} 
    - \frac 1 2 \Delta{\eta^{(2)}_{i,k}}^2\right)
    \label{eq:M}
\end{equation}
where $\widetilde{\eta}_{i,k}^{(1)} =\eta_{i,k}^{(1)} + \Delta \eta_{i,k}^{(1)}$ and $\widetilde{\eta}_{i,k}^{(2)} =\eta_{i,k}^{(2)} + \Delta \eta_{i,k}^{(2)}$ are the random numbers needed to generate the path $\mathbf{x}$ for the $i$th degree of freedom at the target potential. 
The idea\cite{kieninger2021path, X001} is illustrated in Fig.~\ref{fig:OBABO}a, which depicts the sequence of intermediate states during the update from $\bm{x}_k$ to $\bm{x}_{k+1}$.
Since the position update $\bm{q}_k \rightarrow \bm{q}_{k+1}$ ($\mathcal{R}$) depends on the half-step velocity $\dot{\bm{q}}_{k+\frac{1}{2}}$ only, and not on the potential or any random number, both trajectories--under the simulation and the target potential--must pass through the same intermediate states $(\bm{q}_k, \dot{\bm{q}}_{k+\frac{1}{2}})$ and $(\bm{q}_{k+1}, \dot{\bm{q}}_{k+\frac{1}{2}})$.
Under $V(\bm{q})$, the first half-step velocity update $\dot{\bm{q}}_k \rightarrow \dot{\bm{q}}_{k+\frac{1}{2}}$ is obtained by applying the stochastic operator $\mathcal{O}'$ using the random vector $\bm{\eta}_k^{(1)}$, followed by the deterministic update due to the conservative force, $\mathcal{V}'$, evaluated at $\bm{q}_k$. 
Under $\widetilde{V}(\bm{q})$, however, the conservative force update becomes $\widetilde{\mathcal{V}}'$, resulting in a different deterministic contribution. 
To nevertheless reproduce the same half-step velocity $\dot{\bm{q}}_{k+\frac{1}{2}}$, the stochastic operator $\widetilde{\mathcal{O}}'$ must act with a modified random vector $\widetilde{\bm{\eta}}_k^{(1)}$.
An analogous argument applies to the second half-step velocity update $\dot{\bm{q}}_{k+\frac{1}{2}} \rightarrow \dot{\bm{q}}_{k+1}$.
Hence, absolute continuity corresponds to graph congruence in Fig.~\ref{fig:OBABO}a, 
balancing the $\mathcal{O}$ and $\mathcal{V}$ updates.
The difference vectors
$\Delta \bm{\eta}_{k}^{(1)} = \widetilde{\bm{\eta}}_k^{(1)}-\bm{\eta}_{k}^{(1)}$ and
$\Delta \bm{\eta}_{k}^{(2)} = \widetilde{\bm{\eta}}_k^{(2)}-\bm{\eta}_{k}^{(2)}$ are obtained\cite{kieninger2021path, X001} by equating the update operators under simulation potential $V(\bm{q})$ and under the target potential $\widetilde{V}(\bm{q})$ and solving for the random number difference (Appendix~\ref{app:OVRVO}). 
This yields\cite{X001}
\begin{subequations}
\begin{align}
\Delta \eta_{i,k}^{(1)} &= \frac{1}{f'} \frac{\Delta t}{2} \nabla_i U\left(\bm{q}_{k}\right), \\
\Delta \eta_{i,k}^{(2)} &=\frac{d'}{f'} \frac{\Delta t}{2} \nabla_i U\left(\bm{q}_{k+1}\right).
\end{align}%
\label{eq:OVRVO_delta_eta}%
\end{subequations}%
where $\nabla_i U(\bm{q})$ is the perturbation force (Eq.~\ref{eq:V_tilde}) and $d'$ and $f'$ are defined in Eq.~\ref{eq:parameters} using $\Delta t/2$ instead of $\Delta t$ (half a time step).
The framework outlined in sections \ref{sec:pathReweighting} and \ref{sec:GirsanovReweightingFactor} is grounded in the Girsanov theorem,\cite{G182} which links changes in drift to changes in the underlying noise measure. The method is therefore called  \textit{Girsanov reweighting}.

% ----------------------------------------------------------------------------
\section{Implementation Details}
% ----------------------------------------------------------------------------
\label{sec:implementation}

% --- commands for substituting update parameters (full-steps)
\newcommand{\afull}{\Delta t}
\newcommand{\bfull}{-\frac{\nabla_i V(\bm{q})}{m_i}\Delta t}
\newcommand{\dfull}{e^{-\xi\Delta t}}
\newcommand{\ffull}{\sqrt{\frac{k_BT}{m_i}(1-\exp\left({-2\xi\Delta t}\right))}}

% --- commands for substituting update parameters (half-steps)
\newcommand{\aprime}{\frac{\Delta t}{2}}
\newcommand{\bprime}{-\frac{\nabla_i V(\bm{q})}{m_i}\frac{\Delta t}{2}}
\newcommand{\dprimealt}{\exp\left({-\xi\frac{\Delta t}{2}}\right)}
\newcommand{\dprime}{e^{-\xi\frac{\Delta t}{2}}}
\newcommand{\fprime}{\sqrt{\frac{k_BT}{m_i}(1-\exp\left({-\xi\Delta t}\right))}}

% ----------------------------------------------------------------------------
\subsection{The Bussi-Donadio-Parrinello Thermostat}
Our implementation connects to the Bussi-Donadio-Parrinello thermostat\cite{G088,G194} that can be interpreted as a Langevin update scheme \cite{sivak2014time}.
Bussi \textit{et al.} proposed a method for controlling the temperature in a molecular dynamics (MD) simulation based on velocity rescaling, 
% as an extension of the Berendsen thermostat,\cite{berendsen_molecular_1984} 
in order to correctly sample the canonical ensemble.\cite{G088}
In analogy to the  Berendsen thermostat,\cite{berendsen_molecular_1984} the integration algorithm incorporating velocity rescaling is an extension of the velocity-Verlet scheme:\cite{G194,G213}
\begin{subequations}
\begin{align}
\dot{q}_{i,k+1 / 4} & = \alpha\!\left(\dot{\bm{q}}_k, \bm{\eta}^{(1)}_k\right) \, \dot{q}_{i,k},  \\
\dot{q}_{i,k+2 / 4} & = \dot{q}_{i,k+1 / 4} \;-\; \frac{\nabla V(\bm{q}_{k})}{m_i} \frac{\Delta t}{2},   \\
q_{i,k+1} & =q_{i,k} \;+\; \dot{q}_{i,k+2 / 4}\,\Delta t,  \\
\dot{q}_{i,k+3 / 4} & =\dot{q}_{i,k+2 / 4} \;-\; \frac{\nabla V(\bm{q}_{k+1}) }{m_i} \frac{\Delta t}{2}, \\
\dot{q}_{i,k+1} & = \alpha\!\left(\dot{\bm{q}}_{k+3 / 4}, \bm{\eta}^{(2)}_{k}\right) \, \dot{q}_{i,k+3 / 4},
\end{align}%
\label{eq:v-rescale-algo}%
\end{subequations}%
where the scalar velocity rescaling factor {$\alpha\!\left(\dot{\bm{q}}_k, \bm{\eta}^{(z)}_k\right): \mathbb{R}^{3N}\times\mathbb{R}^{3N} \to \mathbb{R}$} encodes the effect of the thermostat. Here, this factor additionally depends on the set of random numbers, $\bm{\eta}^{(z)}_k$, drawn at each integration step, and since it is applied twice, two such sets are required per time step, i.e. $z \in \{1,2\}$.

$\alpha\!\left(\dot{\bm{q}}_k, \bm{\eta}^{(z)}_k\right)$ drives the kinetic energy of the system $K(\dot{\bm{q}_k})$ (Eq.~\ref{eq:E_kin}) towards the kinetic energy at the target temperature $\bar{K} = \frac{1}{2}N_{\! f}k_B T$, where $N_{\! f}$ defines the number of degrees of freedom involved in a thermostatted region ($N_{\! f} = 3N$ in the global scheme).
In their original work, Bussi \textit{et al.}\cite{G088} derive the expression for the stochastic rescaling factor as a full step update.
For the algorithm outlined above,\cite{G213} it must be expressed in \textit{half-steps}, 
\begin{equation}
\alpha\!\left(\dot{\bm{q}}_k, \bm{\eta}^{(z)}_k\right) =  \left[\dfull
+ \frac{\bar{K} \left(1\!-\!\dfull\right)}{K(\dot{\bm{q}}_k)N_{\! f}} \, \left(\sum_{i=1}^{N_{\! f}} {\eta_{i,k}^{(z)}}^2\right)   
+ 2\dprime \sqrt{\frac{\bar{K} \left(1\!-\!\dfull\right)}{K(\dot{\bm{q}}_k)N_{\! f}}} \,\eta_{1,k}^{(z)}  \right]^{\frac 1 2},
\label{eq:alpha_orig}
\end{equation}
under the notational substitutions $R_i \to \eta_i$ and $\tau \to \frac{1}{2\xi}$ compared to Eq.~A7 in Ref.~\citenum{G088}. In the last term, the random number  drawn for only one degree of freedom is used, as shown here with $i=1$.

Eq.~\ref{eq:alpha_orig} defines a velocity rescaling factor that acts uniformly on all $N_{\! f}$ degrees of freedom.
Importantly, the number of degrees of freedom included in this calculation does not need to equal $3N$.
In fact, $N_{\! f}$ can be reduced to define thermalized regions, each governed by its own rescaling factor.
In the limiting case of a \textit{local} (or massive) thermostat, where $N_{\! f}=1$, one finds
\begin{equation}
\sum_{i=1}^{N_{\! f}}{\eta_{i,k}^{(z)}}^2 \;\rightarrow\; {\eta_{i,k}^{(z)}}^2
\qquad\text{and}\qquad
\frac{\bar{K}\left(1\!-\!\dfull\right)}{K(\dot{\bm{q}_k})N_{\! f}} \;\rightarrow\; \frac{k_BT \left(1\!-\!\dfull\right)}{m_i \,\dot{q}_{i,k}^2}.
\end{equation}

Following Eq.~\ref{eq:parameters}, the corresponding \textit{half-step} parameters can be defined using primed symbols,
\begin{align}
    % a'       &= \aprime \cr
    b_i'(\bm{q})  &= \bprime, \cr
    d'       &= \dprimealt, \cr
    f'       &= \fprime,
\label{eq:parameters_half}
\end{align}
which allow Eq.~\ref{eq:alpha_orig} to be rewritten as a rescaling factor defined separately for each degree of freedom:
\begin{align}
\alpha\!\left(\dot{\bm{q}}_k, \bm{\eta}^{(z)}_k\right) 
&\rightarrow \left[ d'^2  + \left(\frac{f'}{\dot{q}_{i,k}} {\eta_{i,k}^{(z)}}\right)^2  + 2d' \frac{f'}{\dot{q}_{i,k}}\eta_{i,k}^{(z)}  \right]^{\frac 1 2} \cr
&= \left| d'  +   \frac{ f'}{\dot{q}_{i,k}}\eta_{i,k}^{(z)} \right|
\cr
&\equiv \left|\alpha_i\!\left(\dot{q}_{i,k}, {\eta}^{(z)}_{i,k}\right)\right|.%
\label{eq:alpha_local}%
\end{align}
The second line in Eq.~\ref{eq:alpha_local} follows from completing the square. 
It is then straightforward to interpret the velocity scaling factor
{$\alpha_i\!\left(\dot{q}_{i,k}, {\eta}^{(z)}_{i,k}\right): \mathbb{R}\times\mathbb{R}\to\mathbb{R}$} as the Ornstein-Uhlenbeck update defined in Eq.~\ref{eq:update_operators}c, here applied over a \textit{half step}.
The algorithm shown in Eq.~\ref{eq:v-rescale-algo} thus corresponds to the $\mathcal{O'V'RV'O'}$ integration scheme,  representing a Langevin integrator with symmetric operator splitting, using  Eq.~\ref{eq:update_operators_R} and 
\begin{subequations}
\begin{align}
\mathcal{V'}\begin{pmatrix} q_{i,k} \\ \dot{q}_{i,k} \end{pmatrix} 
    &= \begin{pmatrix} q_{i,k} \\ \dot{q}_{i,k} + b_i'(\bm{q}_k) \end{pmatrix}, \\
\mathcal{O'}\begin{pmatrix} q_{i,k} \\ \dot{q}_{i,k} \end{pmatrix} 
    &= \begin{pmatrix} q_{i,k} \\ d' \,\dot{q}_{i,k} + f'\, \eta_{i,k} \end{pmatrix},
\end{align}%
\label{eq:update_operators_half_step}%
\end{subequations}%
only that the sign of $\alpha_i$ in Eq.~\ref{eq:alpha_local} is unknown, but can be fixed for $N_{\! f} = 1$ using the condition: 
\begin{equation}
\text{sign}\left(\alpha_i\!\left(\dot{q}_{i,k}, {\eta}^{(z)}_{i,k}\right)\right) = \text{sign}\left(
\eta_{i,k}^{(z)} + \left|d'\frac{\dot{q}_{i,k}}{ f'}\right|
\right),
\label{eq:sign_condition}
\end{equation}
which ensures an integration scheme for underdamped Langevin dynamics.\cite{G194} 
However, in order to get from Eq.~\ref{eq:v-rescale-algo}a  to the $\mathcal{O}'$-step, 
$\alpha_i$ must be multiplied by $\dot{q}_{i,k}$.
Eq.~\ref{eq:sign_condition} assumes that $\dot{q}_{i,k} \ge 0$.
If $\dot{q}_{i,k} < 0$, the upper condition reverses and the chosen sign of $\alpha_i$, following Eq.~\ref{eq:sign_condition}, implies that $\eta_{i,k} \rightarrow -\eta_{i,k}$.
Flipping the sign of the random variable preserves the symmetry of its probability distribution since $p\left(\eta_{i,k}\right) = p\left(-\eta_{i,k}\right)$, which is consistent with Langevin dynamics.
In turn, $\Delta\eta_{i,k}$, required for dynamical reweighting, is inherently unidirectional, so that $p\left(\eta_{i,k} + \Delta\eta_{i,k}\right) \ne p\left(-\eta_{i,k} + \Delta\eta_{i,k}\right)$ (Fig.~\ref{fig:OBABO}b).
Therefore, evaluating the reweighting factor correctly in the Bussi--Donadio-Parrinello scheme requires determining  the sign of $\dot{q}_{i,k}$, ensuring the correct sign of $\eta_{i,k}$ in the scheme $\mathcal{O'V'RV'O'}$.

% For a \textit{global thermostat}, evaluating the reweighting factor in Eq.~\ref{eq:alpha_orig} leads to a system of coupled equations of motion involving all degrees of freedom. 
% As a consequence, any redistribution of force contributions between the $U(\bm{q})$-dependent and $\bm{\eta}$-dependent terms for a single coordinate affects the entire phase space. 
% This induces transitions under $V(\bm{q})$ that cannot occur under $\widetilde{V}(\bm{q})$, violating absolute continuity. 
% Accordingly, Girsanov reweighting is not applicable in the global scheme.

% ----------------------------------------------------------------------------
\subsection{The CP2K Code}
\begin{figure}
% ---------------- left block (a) ----------------
\begin{minipage}[t]{0.64\linewidth}
a)\hfill\vspace{2mm}

\footnotesize
% \begin{minted}{fortran}
\begin{verbatim}
TYPE degree_of_freedom_type
  REAL(KIND=dp), DIMENSION(2) :: grad_U = 0.0_dp
  REAL(KIND=dp), DIMENSION(2) :: eta = 0.0_dp
  REAL(KIND=dp), DIMENSION(2) :: delta_eta = 0.0_dp
  REAL(KIND=dp)               :: mass
  REAL(KIND=dp)               :: vel_sign = 0.0_dp
END TYPE

TYPE girsanov_env_type
  INTEGER                     :: mode, n_atoms, n_dim
  INTEGER, POINTER, DIMENSION(:) :: atom_list => NULL()
  INTEGER, POINTER, DIMENSION(:, :) :: index_map => NULL()
  INTEGER, POINTER, DIMENSION(:) :: thermo_map => NULL()
  REAL(KIND=dp)               :: U = 0.0_dp
  REAL(KIND=dp)               :: delta_S = 0.0_dp
  REAL(KIND=dp)               :: xi, prefactor_eta_1, &
                                 prefactor_eta_2, &
                                 temperature, dt, kBT
  TYPE(degree_of_freedom_type), DIMENSION(:), POINTER :: &
                               degrees_of_freedom => NULL()
END TYPE
\end{verbatim}
% \end{minted}
\end{minipage}
\hfill
%
% --------- vertical separator line ---------
\vrule width 0.7pt
%
% ---------------- right block (b) ----------------
\hfill
\begin{minipage}[t]{0.35\linewidth}
\hspace{0.5cm}b)\hfill\vspace{2mm}

\footnotesize
\begin{verbatim}
    &GIRSANOV
      MODE BIAS
      LOG_M {real}
      &BIAS
        SOURCE METADYNAMICS |\
        EXTERNAL_POTENTIAL |\
        RESTRAINT
        ATOM_LIST {integer} ..\
                   {integer}
      &END BIAS    
      &PRINT
        &GIRSANOV_INFO
        &END GIRSANOV_INFO
        &REWEIGHTING_FACTOR
        &END REWEIGHTING_FACTOR
        &RANDOM_NUMBERS
        &END RANDOM_NUMBERS		  
        &BIAS_FORCE
        &END BIAS_FORCE				
      &END PRINT        
    &END GIRSANOV          
\end{verbatim}
\end{minipage}
\caption{a) The two newly introduced Fortran types that travel through the MD cycle to gather the necessary data for calculating the reweighing factor.
b) New Girsanov section as \texttt{MD} subsection. Each \texttt{PRINT} section represents a standard CP2K \texttt{PRINT} accepting keywords such as \texttt{FILENAME} and the \texttt{EACH} subsection for defining the output rate.}
\label{fig:fortran_types-grw_input}
\end{figure}

CP2K provides multiple implementations of Eq.~\ref{eq:ULE}. We have incorporated our implementation into the native velocity-Verlet integrator, combined with the CSVR (canonical sampling through velocity rescaling) thermostat, which realises the Bussi–Donadio–Parrinello scheme.\cite{G088,G194}
In the context of biased MD simulations, the perturbation potential $U(\bm{q})$ is formally expressed as a bias potential $U_{\text{bias}}(\bm{q}) = -U(\bm{q})$ applied to $\widetilde{V}(\bm{q})$ and a biasing force $\nabla{}U_{\text{bias}}(\bm{q}) = -\nabla{}U(\bm{q})$.\cite{M076}

CP2K is primarily implemented in Fortran 2008 and achieves high performance on parallel architectures through hybrid parallelisation using MPI, OpenMP, and CUDA.
Its modules are object-based, using derived types and encapsulation for clarity.
With the implementation of Girsanov reweighting, two new Fortran objects (\texttt{TYPE}, Fig.~\ref{fig:fortran_types-grw_input}a) 
form the basis for tracking the integrator during the MD run within CP2K: A global Girsanov environment, and a sub-environment for each monitored degree of freedom.
The central idea is that these objects are passed through the MD loop, collecting the data required for calculating the reweighting factor. This includes crucial information such as the two random numbers generated at each time step, the bias potential and biasing force applied to the system, the velocity sign, the coupling strength of the thermostat, the temperature, the particle mass, and the integration time step. In addition, the objects acquire information on how the degrees of freedom are distributed across MPI processes and how these map onto the Girsanov-internal list of biased degrees of freedom.

Figure~\ref{fig:fortran_types-grw_input}b illustrates the new input section that has been added to the \texttt{MD} section of CP2K. This section defines the source of the bias and offers three options: metadynamics, external potentials, and restraint potentials. The latter two are native features of CP2K, whereas for metadynamics, the user can choose between the native implementation and the PLUMED interface, which provides full PLUMED functionality (beyond metadynamics itself).\cite{bonomi_promoting_2019}
Subsequently, the input specifies the atoms to be monitored by the Girsanov environment, where a value of $-1$ indicates that all degrees of freedom are considered.
It is planned in the future to add a perturbation mode, where the biasing force is calculated but not applied to the system (e.g., after a linear response calculation) as well as mixing of \texttt{FORCE\_EVAL} sections, offered by CP2K.

Finally, an extensive \texttt{PRINT} subsection sets the output options, enabling users to obtain the static and dynamic reweighting factors, the biasing forces and the random numbers generated at each time step.
These output sections are in accordance with CP2K standard output conventions and support customized file names, adjustable output frequencies (via the \texttt{EACH} subsection), and flexible organization of iteration levels for file generation.
It should be noted that outputting random numbers and biasing force for the ex-post calculation  of the reweighting factor is only meaningful when the output frequency is set to $1$ (i.e., at every time step); the Girsanov modules issue a warning if this is not the case.
Accordingly, the Girsanov section can produce up to three distinct output files using the extensions
\texttt{.girsanov\_eta}, \texttt{.girsanov\_bias}, and \texttt{.girsanov\_factor}.
The first two files contain the raw information obtained from the MD integrator that is used to calculate the reweighting factors. 
For each frame, two header lines are printed: the total number of monitored degrees of freedom, and the frame info together with the current value of $U(\bm{q})$.
This is followed by the entries for each monitored degree of freedom (according to \texttt{ATOM\_LIST}).
For \texttt{.girsanov\_eta}, these lines contain particle index, dimension index, $\eta_i^{(1)}$, $\eta_i^{(2)}$, $\Delta\eta_i^{(1)}$, and $\Delta\eta_i^{(2)}$.
In \texttt{.girsanov\_bias}, one gets particle index, dimension index, particle mass, $\nabla_i{U(\bm{q})}$.
For non-debugging runs, the file ending with \texttt{.girsanov\_factor} (invoked via \texttt{REWEIGHTING\_FACTOR}) is typically sufficient, containing both static and dynamic reweighting factors at each output step in their logarithmic form. 
It follows the approach of Schäfer \textit{et al.}\cite{M076}, with
\begin{equation}
    \texttt{log(g)} := -\ln\left(\tfrac{\widetilde{Z}}{Z}g\right) = \frac{U(\mathbf{q}_k)}{k_BT}
\end{equation}
and 
\begin{equation}
     \texttt{log(M)} := -\ln\left(M[ \mathbf{x} | \bm{x}_0 ]\right) = \sum_{k=0}^{n_{\text{interval}}-1} \frac{\Delta S_{k}}{D},
\label{eq:logM}
\end{equation}
where 
\begin{align}
\frac{\Delta S_{k}}{D} = \sum_{k'=0}^{n_{\text{out}}-1}\sum_{i=1}^{3N}
& \biggl(\left.\eta^{(1)}_{i,k n_{\text{out}}+k'}\cdot\Delta\eta^{(1)}_{i,k n_{\text{out}}+k'} \right.  \left. + \frac 1 2 \Delta{\eta^{(1)}_{i,k' n_{\text{out}}+k'}}^2 \right. \cr
& \left. + \eta^{(2)}_{i,k n_{\text{out}}+k'}\cdot\Delta\eta^{(2)}_{i,k n_{\text{out}}+k'}  \right. \left. + \frac 1 2 \Delta{\eta^{(2)}_{i,k n_{\text{out}}+k'}}^2 \right).
\label{eq:buffer}
\end{align}
$n_{\text{interval}}$ is the length of the sampled path and $n_{\text{out}}$ the output rate defined through the \texttt{PRINT} section. 
It should be noted that for \texttt{log(g)} only stationary points are required.
Hence, \texttt{REWEIGHTING\_FACTOR} produces a list of values for \texttt{log(g)} and \texttt{log(M)} at every output step $k$.
According to Eq.~\ref{eq:buffer}, \texttt{log(M)} accumulates the dynamical reweighting factor over time steps $k$ according to the \texttt{PRINT} frequency and is set to zero after $n_{\text{out}}$ steps.
If the output cycle is interrupted prematurely, the instantaneous value of \texttt{log(M)}  must therefore be stored in the restart file. 
This is achieved using the real-valued keyword \texttt{LOG\_M}, which has a default value of zero.
If PLUMED is used, it must be ensured that the RESTART keyword is set also in the PLUMED intput file.

\begin{figure*}[ht]
    \centering
    \includegraphics[width=\linewidth]{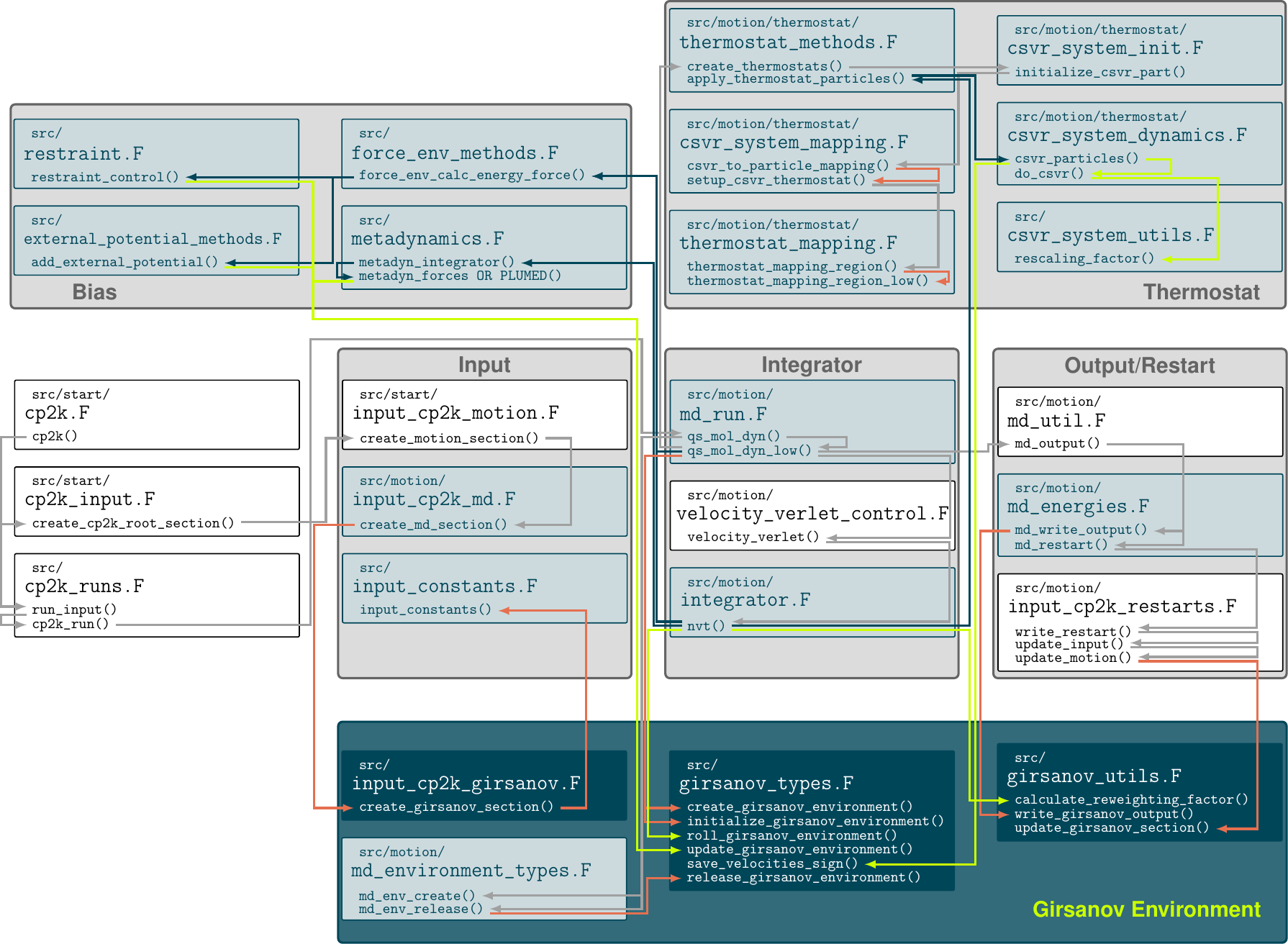}
    \caption{Call diagramm of all CP2K modules involved in Girsanov reweighting using the CSVR thermostat. Each box refers to the file name and \texttt{MODULE} that can be found under the source path given in the upper line. \texttt{SUBROUTINES} are listed below the \texttt{MODULE} name and arrows indicate the \texttt{CALL} command.
    Light blue boxes correspond to \textit{existing} modules that have been changed, whereas dark blue boxes correspond to \textit{new} modules. Calls for setting up, configuring, and releasing the Girsanov environment are marked in orange. Runtime calls updating the Girsanov environment within the MD loop are shown in yellow.}
    \label{fig:cp2k-code}
\end{figure*}

When implementing Girsanov reweighting, our goal was to encapsulate the new code as much as possible in line with CP2K standards. However, the new Girsanov environment needs to collect information throughout the MD loop, making slight adjustments throughout the entire MD machine unavoidable. 
Figure~\ref{fig:cp2k-code} presents the complete network of \texttt{MODULES} and \texttt{SUBROUTINES} involved in an MD run using Girsanov reweighting, illustrated as a \textit{call diagram}.
In addition to 16 updated modules, three new modules form the Girsanov capsule.
The creation of the Girsanov input section is the first procedure executed, regardless of whether Girsanov reweighting is ultimately used or not.
The programme then proceeds to the MD modules (naturally, \texttt{RUN\_TYPE} must be set to MD) and requires the NVT ensemble with a MASSIVE CSVR thermostat when Girsanov reweighting is enabled.

In many cases, the Girsanov environment is simply passed through the modules as an \texttt{OPTIONAL} variable until it reaches its final destination (e.g., the subroutine \texttt{rescaling\_factor}, which draws the random number).
However, the distribution of the configuration space over the MPI processes as well as the running thermostats require careful mapping between the Girsanov and the MD environment to avoid losing information or, worse, double counting it.
For this reason, the Girsanov environment already chaperones the thermostat creation process and retrieves the \texttt{thermo\_map}, going deep into the structure of the CSVR thermostat. There, it also collects valuable information on the thermostat settings (notably the coupling constant).

After the creation and initialisation of the required objects (MD environment, force environment, Girsanov environment, etc.), the programme enters the MD loop and calls the velocity-Verlet integrator for the NVT ensemble, which implements Eq.~\ref{eq:v-rescale-algo}. 
The Girsanov environment is passed along to monitor both the force calculation and the thermostat.
Obtaining the bias potential and forces from the former is straightforward, since in all three supported cases (metadynamics, external potential, and restraint potential) these contributions are explicitly applied to the system and subsequently recorded by the Girsanov environment.
However, saving the random number, is more complicated, as it is less exposed and requires access to the lowest level of iteration where the CSVR rescaling factor is computed (ensuring that $N_f = 1$).
In addition, while the new implementation benefits from native MPI broadcasts of the force environment, this is not the case for random numbers, which therefore have to be explicitly called within the Girsanov module.

At the end of the MD step, the Girsanov reweighting factors are calculated and stored in the buffer for \texttt{log({M})}, depending on the output rate $n_{\text{out}}$ in Eq.~\ref{eq:buffer}, or written to file.
For each degree of freedom, the Girsanov environment is then "rolled", which is almost like a reinitialisation process, except that $\nabla U(\mathbf{q}_{k+1}) \rightarrow \nabla U(\mathbf{q}_k)$.
Upon completion of the MD run, all dynamically allocated memory must be properly deallocated to ensure a clean termination of the simulation. This triggers a hierarchical sequence of \texttt{release} calls propagating through the respective parent modules. Within this sequence, the Girsanov environment is explicitly released by the MD environment.

The method introduces only minimal additional computational overhead. Even in simulation settings where the biasing force must be computed additionally (only required in the perturbation mode), this calculation is low-dimensional with negligible cost compared to the $3N$ force evaluations.
The reweighting factor is evaluated once per MD-step by looping over all degrees of freedom involving only elementary arithmetic operations. Consequently, the overall scaling with system size remains linear, and the additional cost is negligible compared to force evaluation and thermostat propagation. 
A massive thermostat is generally more computationally demanding than a global one.\cite{M076}

% ----------------------------------------------------------------------------
\section{Validation}
% ----------------------------------------------------------------------------
\label{sec:observables}

% ----------------------------------------------------------------------------
\subsection{Time Correlation Functions}

In this work, dynamical properties are evaluated using TCFs, to illustrate the performance of the implementation.
In linear response theory, TCFs quantify the time-lagged response of a system to weak external perturbations, yielding transport coefficients via the Green–Kubo formalism, or spectral density functions.\cite{berne_time-dependent_1971, kubo1966fluctuation, B048, B027}
Furthermore, TCFs can be used to characterise the intrinsic dynamics of a system at equilibrium, such as transitions between discrete states over time, using Markov state models (MSMs).\cite{swope2004describing, buchete2008coarse, keller2010comparing, prinz2011markov}

The TCF of two observables $a$ and $b$ can be reformulated in terms of the path probability density, following Eq.~\ref{eq:path_measure}, where the lag time $\tau$ is determined by the discretised path length, $\tau = n\cdot\Delta{}t$,\cite{R086}
\begin{align}
 C_{ab}(\tau) &= \langle a(0) b(\tau) \rangle \cr
 &=\int_\Gamma \int_\Gamma \d \bm{x}_0\,\d \bm{x}_n\; p(\bm{x}_0)a(\bm{x}_0) \,  p(\bm{x}_0, \bm{x}_n; \tau)b(\bm{x}_n)  \cr
 &=\int_\Gamma \d \bm{x}_0 \; p(\bm{x}_0)a(\bm{x}_0) \int_{\mathcal{S}}\mathcal{D}\mathbf{x} P[\mathbf{x}] b(\bm{x}_n),
 \label{eq:tcf_path}
\end{align}
where $a(t) \equiv a(\bm{x}(t))$.
$p(\bm{x}_0, \bm{x}_n; \tau)$ denotes the conditional probability of reaching state $\bm{x}_n$ starting from state $\bm{x}_0$ through all possible paths of length $\tau$.
Accordingly, the reweighted TCF can be obtained by insertion of Eq.~\ref{eq:reweighted_expected_path}:
\begin{align}
 \widetilde{C}_{ab}(\tau) 
 =&\int_\Gamma \d \bm{x}_0 \; g(\bm{x}_0)p(\bm{x}_0)a(\bm{x}_0) \int_{\mathcal{S}}\mathcal{D}\mathbf{x} M[\mathbf{x}] P[\mathbf{x}] b(\bm{x}_n)\cr
 \equiv& \;  \langle g(0)a(0) M(\tau)b(\tau) \rangle,
 \label{eq:tcf_path_rw}
\end{align}
where the last line introduces a shorthand notation of the reweighted TCF.

\subsection{Conditional Path Weights and Reweighting Stability}
If the path observable $s$ is set to unity, Eq.~\ref{eq:reweighted_expected_path} yields the average probabability of finding the simulated paths of length $\tau$ at the target potential $\widetilde{V}(\bm{q})$ (conditional path weight). 
This is equivalent to calculating the average over $\mathcal{S}$ of the dynamical reweighting factor $M(\tau)$, given as 
\begin{equation}
    \langle M(\tau)\rangle = \lim_{|A| \rightarrow \infty}  \frac{1}{|A|}\, \sum_{\mathbf{x}\in A} M(\mathbf{x}|\mathbf{x}_0)
    = \lim_{|A| \rightarrow \infty}  \frac{1}{|A|}\, \sum_{\mathbf{x}\in A}\prod_{k=0}^{n_{\tau}} \exp\left(-\frac{\Delta S_k}{D}\right).
    \label{eq:M_average}
\end{equation}
These path weights can also be interpreted as a stability measure of the dynamical reweighting. If the value $\langle M(\tau)\rangle$ decays, the probability of finding the simulated path with $\widetilde{V}(\bm{q})$ decreases.
For $\langle M(\tau)\rangle = 0$, absolute continuity is no longer satisfied. 
Hence, efficient Girsanov reweighting can only be carried out, if $\langle M(\tau)\rangle$ remains close to unity, the time window of which will be referred to as \textit{stable region} for reweighting.

% ----------------------------------------------------------------------------
\subsection{Dynamical Properties}

\subsubsection{Markov State Model (MSM)}
Let the configuration space be discretised into $n$ disjoint states $\chi_n \in \Omega \subset \Gamma$, where $p_i(t)$ denotes the probability of finding the system in state $\chi_i$ at time $t$.
For a Markovian, reversible, and ergodic process, the transfer operator $\mathcal{T}_{\tau}$ describing the time evolution of the (true) probability density can thus be approximated as a transition matrix $\mathbf{T}(\tau)$ with dimension $n\times n$ acting on $\bm{p}(t) = \big(p_1(t), p_2(t), ..., p_n(t)\big)$,
\begin{equation}
    \bm{p}(t+\tau) = \mathbf{T}(\tau)\bm{p}(t),
\end{equation}
where $\tau$ denotes the lag time \cite{prinz2011markov}. 
The dominant eigenvalues $\lambda_k(\tau)$ and (left) eigenfunctions $\bm{l}_k(\mathbf{x})$ of $\mathbf{T}(\tau)$, obtained from the equation,
\begin{equation}
    \bm{l}_k(\bm{x})\mathbf{T}(\tau)  = \lambda_k(\tau)\bm{l}_k,  
\label{eq:MSM_propagator_eigenvalues}
\end{equation}
provide important insight into the microscopic dynamics of the system, including the identification of long-lived metastable states and the transition rates between them.
The corresponding slow dynamical processes are quantified by the implied time scales,
\begin{equation}
    t_k^{\text{its}} = - \frac{\tau}{\ln\!\left(\lambda_k(\tau)\right)},
\end{equation}
where the approximation is considered to be valid if the $t_k^{\text{its}}$ are independent of $\tau$.

$\mathbf{T}(\tau)$ can be estimated from MD simulations using the transition probability $C_{ij}(\tau)$ from state $\chi_i$ to state $\chi_j$ after $\tau$,
\begin{equation}
    T_{ij}(\tau) = \frac{C_{ij}(\tau)}{\sum_j{C_{ij}(\tau)}}.
\end{equation}
$C_{ij}(\tau)$ is expressed as time-correlation function, following Eq.~\ref{eq:tcf_path},
\begin{equation}
C_{ij}(\tau) = \left\langle1_{\chi_i}(0)\, 1_{\chi_j}(\tau)\right\rangle,
\label{eq:C_from_TCF}
\end{equation}
where the indicator function $1_{\chi_i}$ is unity only if the system is in state $\chi_i$, and zero elsewhere.

Eq.~\ref{eq:C_from_TCF} can be reweighted according to Eq.~\ref{eq:tcf_path_rw}, where the factor $\frac{Z}{\widetilde{Z}}$ obtained for $g(\bm{x_0})$ in Eq.~\ref{eq:g} cancels upon normalisation of $\mathbf{T(\tau)}$.

\subsubsection{Self-Diffusion Coefficient}

Fick's second law of diffusion relates the change in concentration $c$ of a particle over time with its Laplacian,\cite{fick_ueber_1855}
\begin{equation}
    \frac{\partial c}{\partial t}= D \Delta c.  
\end{equation}
The self-diffusion coefficient, represented by $D$, has units of area per time.
From an MD simulation, it can be obtained from the time integral of the velocity autocorrelation function (VACF) of the diffusing particles,\cite{B048,G193,kubo1966fluctuation}
\begin{equation}
    D= \frac{1}{3N}\int_{0}^{\infty} \sum_{i=1}^{3N}\langle \dot{q}_i(0) \dot{q}_i(\tau) \rangle  \,\mathrm{d}\tau  ,
\label{eq:GK-diffusion}
\end{equation}
The factor $\frac{1}{3N}$ accounts for the fact that $N$ particles diffuse in three dimensions.

Eq.~\ref{eq:GK-diffusion} can be reweighted according to Eq.~\ref{eq:tcf_path_rw}.
To obtain a valid probability measure, the Girsanov weight must be normalised such that its expectation equals unity.
Although this requirement is formally related to the partition-function ratio 
$\frac{Z}{\widetilde{Z}}$ (cf. Eq.~\ref{eq:g}), in practice it is ensured directly by dividing by the mean path weight.
This yields the estimator of the reweighted diffusion coefficient,
\begin{equation}
\widetilde{D} = \frac{1}{3N}\int_{0}^{\infty} \sum_{i=1}^{3N} \frac{\langle g(0)\dot{q}_i(0) M(\tau)\dot{q}_i(\tau) \rangle}{\langle g(0) M(\tau)\rangle} \,\mathrm{d}\tau .
\end{equation}

% ----------------------------------------------------------------------------
\subsection{Computational Details}
The simulations were performed using a development version of CP2K (commit \texttt{983d0ee}, derived from commit \texttt{2f51be0} on the \texttt{master} branch; available on GitHub\cite{CP2K,GitHub_CP2K,GitHub_CP2K_Girsanov}). The PLUMED version used was 2.9.2.\cite{tribello_plumed_2014,bonomi_promoting_2019}
Python~3.13.5. was used for post processing and plotting using the 
following libraries:
NumPy~2.3.1, SciPy~1.15.3, ChirPy~0.30.2, and Matplotlib~3.10.3.

% full hash 
% 117c2eb0f5e0a23ed7c70b40075beafdcabc3a75
% and 
% 2f51be0cd664684a312da5e722ab7f98bfce3a24

\subsubsection{Simple Rerun Benchmarks}
All implemented sources of biasing forces accessible within the new Girsanov environment were benchmarked using strong bias potentials to generate reference trajectories, setting $\widetilde{V}(\bm{q}) = 0$.
The values of $\Delta\eta^{(z)}$, computed with CP2K, were then validated by rerunning the trajectories from the same starting point with $\eta^{(z)} +\Delta\eta^{(z)}$ \underline{instead} of applying $U_{\text{bias}}(\bm{q})$, using a lightweight $\mathcal{O'V'RV'O'}$-implementation in Python (see the supporting material linked in the Data Availability statement for the CP2K input files and rerun scripts).

A single Lennard-Jones (LJ) particle with parameters $\epsilon=119.8\,k_B$, $\sigma=3.405$~Å, and a mass of 39.948~amu (argon) was placed in a cubic $20\times20\times20$~Å$^{3}$ cell with periodic boundary conditions. 
The temperature was set to $T=100$~K and a massive CSVR thermostat with a coupling constant of $\tau=1$~fs ($\xi=500$~ps$^{-1}$) was used.
The systems were propagated with a time step of $\Delta=5$~fs.

To generate a circular trajectory, the starting position was set to $\bm{q}_0 = (5, 0, 10)$~Å.
A radial harmonic restraint of the form $U_{\text{bias}}(r) = \frac{1}{2}k_r(r-a)^2$ was applied, where $r = \sqrt{x^2 + y^2}$, using PLUMED  (through the \texttt{\&METADYNAMICS} section).
The target radius was set to $a=5$~Å with a radial force constant of $k_r=1000$~kJ mol$^{-1}$ Å$^{-2}$.
Circular motion was enforced by applying a moving harmonic restraint of the form $U_{\text{bias}}(\theta, t) = \frac{1}{2}k_\theta(\theta-\alpha(t))^2$ to the angular coordinate $\theta = \arctan\!2(y,x)$ with a force constant of $k_\theta=10000$~kJ mol$^{-1}$ rad$^{-2}$.
The angular restraint was advanced over 15~ps (3000 steps) to span the full range of $\alpha$ from 0 to $2\pi$, so that $\alpha(t) = \frac{2}{15}\pi t$ where the simulation time $t$ is measured in ps.

To generate a linear trajectory, the starting position was set to $\bm{q}_0 = (10, 10, 10)$~Å.
A linear bias potential of the form $U_{\text{bias}}(x) = m x$ was applied with slope $m = 20$~kJ mol$^{-1}$ Å$^{-1}$ using PLUMED (through the \texttt{\&METADYNAMICS} section) or the \texttt{\&EXTERNAL\_POTENTIAL} section of CP2K.
The system was run for 1~ns (200,000 steps).

To impose a potential well using the \texttt{\&CONSTRAINT} section of CP2K, a fixed point was defined as potential minimum at $\bm{O} = (1, 1, 1)$~Å.
The starting position of the particle was set to $\bm{q}_0 = (10, 10, 10)$~Å.
A harmonic restraint $U_{\text{bias}}(\bm{q}) = k (\bm{q} - \bm{O})^2$ was applied with a force constant of $k=10$~kJ mol$^{-1}$ Å$^{-2}$.
The system was run for 50~ps (10000 steps).

% ----------------------------------------------------------------------------
\subsubsection{Markov State Model: 1D-Periodic Double Basin}
To emulate a 1D double well potential within the 3D framework of CP2K, a periodic double basin potential of the form 
$$
\widetilde{V}(\bm{q}) = 
\frac{1}{4} \Biggl[ 
k_L\left(1-\sin(2x)\right) + 
k_R\left(1+\sin(2x)\right)  \Biggr]  \left(y^2 + z^2\right) 
+ a \cos^2(2x)
$$
has been defined through the \texttt{\&EXTERNAL\_POTENTIAL} section in a tetragonal $\pi\times10\times10$~rad Å$^{2}$ cell using periodic boundary conditions, with
radial harmonic constraints for the left and right well, 
$k_L = 13795$~kJ mol$^{-1}$ Å$^{-2}$ and 
$k_R = 10900$~kJ mol$^{-1}$ Å$^{-2}$, respectively, as well as
a barrier height $a = 7.74$~kJ mol$^{-1}$.
The $x$-dimension represents the periodic double well potential, while a two-dimensional radial plane is defined by $y$ and $z$, with $r = \sqrt{y^2 + z^2}$.

Ghost particles with zero charge, zero radius, and mass 39.948~amu were randomly placed on the potential,
while $x$ was sampled uniformly from the interval $[0, \pi)$, 
while $y$ and $z$ were drawn randomly from narrow uniform distributions centered at zero in the interval $[-0.015, 0.015]$~Å  to place the particles close to the $x$-axis.
The temperature was set to $T=300$~K and a massive CSVR thermostat with a coupling constant of $\tau=100$~fs ($\xi=5$~ps$^{-1}$) was used.
The systems were propagated with a time step of $\Delta=0.5$~fs.
The output time step for the trajectory and reweighting factors was 1~fs (every second step).

For the unbiased simulation, 100 particles were considered and the system was equilibrated for 5~ps, followed by a production run of 1~ns. 
As there is no interaction between ghost particles, this setup is equivalent to 100 independent runs using a single particle.
For the biased, a bias potential of the form 
$U_{\text{bias}} = \sin^2(2x)$ was applied using PLUMED (through the \texttt{\&METADYNAMICS} section).
60 starting configurations were drawn from the unbiased production run after 0.5~ns and individually run for 1~ns each. 

Another set of simulations using the same setup were carried out with a particle mass of 12.01~amu and a thermostat coupling constant of $\tau=50$~fs ($\xi=10$~ps$^{-1}$), using 30 starting configurations.

For subsequent analysis, the same number of particles (60 or 30) was used for both the unbiased and biased simulations. 
The MSM was constructed with Deeptime~0.4.5, which provides built-in functionality for Girsanov reweighting.\cite{M076}
A grid of 51 states was generated by discretising the $x$-dimension over the interval $0 < x \le \pi$.
The lag times were sampled uniformly within the following ranges:
$0.01 \le \tau \le 0.2$ (10 values),
$0.25 \le \tau \le 0.7$ (10 values), and
$1 \le \tau \le 10$ (15 values).

Input files and python scripts for postprocessing are provided in the supporting material (see Data Availability statement).

% ----------------------------------------------------------------------------
\subsubsection{Diffusion in a Lennard-Jones Liquid}
108 LJ particles with parameters $\epsilon=119.8\,k_B$, $\sigma=3.405$~Å, and a mass of 39.948~amu (argon) were randomly placed in a cubic $17.158\times17.158\times17.158$~Å$^{3}$ cell with periodic boundary conditions, ensuring a minimal initial distance of $\ge3$~Å. 
This corresponds to a density of 1.4184~g~cm$^{-1}$.
The cutoff for the LJ potential was set to $r_{\text{cut}}=8.4$~Å.
The temperature was set to $T=85$~K and a massive CSVR thermostat with a coupling constant of $\tau=50$~fs ($\xi=10$~ps$^{-1}$) was used.
The systems were propagated with a time step of $\Delta=5$~fs.
The output time step for the trajectory and reweighting factors was 5~fs (every step).

The unbiased and biased simulations were carried out within the same configuration through biasing 10 (out of 108) particles with a linear bias potential 
$U_{\text{bias}}(x) = m x$ with slope $m = 2$~kJ mol$^{-1}$ Å$^{-1}$, using the \texttt{\&EXTERNAL\_POTENTIAL} section of CP2K. 
With this, we approximately considered the forces acting on the biased particles as independent, so that a single biased simulation effectively corresponded to 10 independent runs.
The system was first equilibrated for 15~ps, followed by a 1~ns production run. 
A total of 10 simulations were conducted from the same initial configuration, each time selecting a different set of 10 biased particles (100 biased particles in total).

For subsequent analysis, the same number of particles (100) was used for both the unbiased and biased simulations. 
For each simulation, the first 50 ps were excluded from the analysis.
The VACF was calculated with ChirPy and NumPy using non-overlapping time windows with a length of 25~ps (18 windows per particle).
For Girsanov reweighting, the increments $\frac{\Delta S}{D}$ and $\frac{U}{RT}$ were calculated analytically for each particle based on the values for $\Delta\eta^{(z)}$ provided by CP2K, and validated against the CP2K output which provides only one summed reweighting factor for all biased degrees of freedom.
The VACF was evaluated on non-overlapping time windows of length 25~ps.
The computation was implemented in a fully vectorised manner using NumPy, thereby improving numerical efficiency.
The time-resolved diffusion constant was obtained as the cumulative sum of the individual VACFs. 

Input files and python scripts for postprocessing are provided in the supporting material (see Data Availability statement).

% ----------------------------------------------------------------------------
% \subsubsection{Markov State Model of (\textit{R})-Butan-2-ol}
% One molecule in the gas phase at 300 and 500~K.

% Bias on  C-C-C-C 

% Maybe 2D-MSM: colvar1 = C-C-C-C, colvar2 = H-O-C-C

% \newpage
% ----------------------------------------------------------------------------
\section{\label{sec:Results}Results and Discussion}
% ----------------------------------------------------------------------------

% ----------------------------------------------------------------------------
\subsubsection{Simple Rerun Benchmarks}
\begin{figure}[t]
    \centering
    \includegraphics[width=\textwidth]{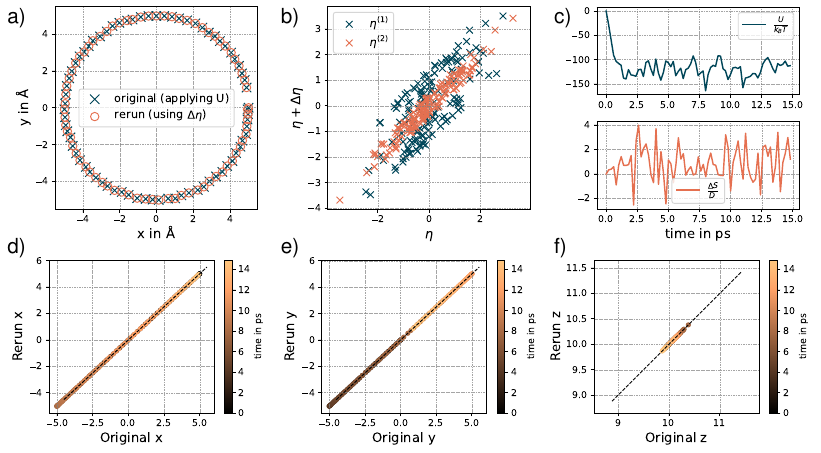}
    \caption{Simple Rerun Benchmarks: Results of the rerun benchmark of a Lennard-Jones particle forced onto a circular trajectory using CP2K and PLUMED.
    a) Superposition of the trajectories generated with CP2K applying the bias $-U(\bm{q})$ and the ones generated with the $\mathcal{O'V'RV'O'}$ script using the $\Delta\eta^{(z)}$ provided by the CP2K run;
    b) Correlation of $\eta^{(z)}$ and $\widetilde{\eta}^{(z)}=\eta^{(z)} + \Delta\eta^{(z)}$;
    c) Static (blue) and dynamic (orange) reweighting factor increments in their logarithmic form;
    d-f) Correlation of the positions in $x$, $y$, $z$, respectively, between the original run and the rerun. The color indicates the time step (black: starting point).}
    \label{fig:results_rerun_PLUMED_c}
\end{figure}

Fig.~\ref{fig:results_rerun_PLUMED_c} collects the results of the rerun benchmarks using PLUMED that has been accessed through the \texttt{\&METADYNAMICS} section of CP2K
(i.e. set \texttt{SOURCE METADYNAMICS} in the Girsanov input) to force an LJ particle on a circular trajectory. 
Further results for other bias sources are collected in Figs.~\ref{fig:results_rerun_PLUMED_l}-\ref{fig:results_rerun_restraint} in Appendix~\ref{app:results}.

All benchmarks show excellent agreement between the trajectory generated with CP2K applying the bias $-U(\bm{q})$ and the ones generated with the $\mathcal{O'V'RV'O'}$ script using the $\Delta\bm\eta^{(z)}$ provided by the CP2K run (Fig.~\ref{fig:results_rerun_PLUMED_c}a).
This allows us to verify that collecting the bias force through the Girsanov environment, as well as the subsequent calculation of $\Delta\bm\eta^{(z)}$ is correctly implemented.
To truly see and benchmark the effect of the bias on the trajectory, a very strong potential had to be applied. 
The required change of $\bm\eta^{(z)}$, in order to compensate for the bias is shown in Fig.~\ref{fig:results_rerun_PLUMED_c}b.
While the absolute value of the change $\Delta\bm\eta^{(1)}$ ranges up to two, 
$\Delta\bm\eta^{(2)}$ remains much smaller, which is a natural consequence of the scaling factor $\d' \le 1$ (Eq.~\ref{eq:OVRVO_delta_eta}b).
Given the strength of both the static and the dynamic reweighting factors (see Fig.~\ref{fig:results_rerun_PLUMED_c}c and d), the required changes in $\bm\eta^{(z)}$ are small with respect to the full range of the normal distribution.
For the other rerun benchmarks, shown in Appendix~\ref{app:results}, the required change in $\bm\eta^{(z)}$ is even smaller, despite the similarly large bias potential.
It should be noted that $\Delta\bm\eta^{(z)}$ is determined by the gradient $\nabla U(\bm{q}_k)$ alone and not by the magnitude of $U(\bm{q}_k)$.
In the case of a circular trajectory, as shown in Fig.~\ref{fig:results_rerun_PLUMED_c}, 
$\nabla U(\bm{q}_k)$ is nourished by the centripetal force, which turns out large as it has to compensate for the centrifugal force--an effect absent in the other rerun benchmarks.

The rerun trajectories remain numerically stable until the end of the maximum observation window of 200,000 time steps (1~ns) (see Figs.~\ref{fig:results_rerun_PLUMED_l} and~\ref{fig:results_rerun_external}~d-f, respectively).
This is sufficient for all reweighting scenarios in atomistic simulations, since the lag times of the slowest processes typically lie below this value.

All rerun benchmarks were carried out at a finite temperature, so all trajectories appear rugged due to particle inertia. 
One might inquire whether it would be possible to observe the "pure" effect by applying a bias force to a particle at rest (i.e. at 0~K). 
However, examining the definition of $f'$ in Eq.~\ref{eq:parameters_half}, it becomes clear that the (target) temperature of the ensemble is required to connect the $\bm{q}$ domain with the $\bm{\eta}$ domain. 
Consequently, Girsanov reweighting is only defined at a finite temperature, which is a direct consequence of the formalisation of Langevin dynamics and the canonical ensemble.

% ----------------------------------------------------------------------------
\subsubsection{Markov State Model: 1D-Periodic Double Basin}
\begin{figure}[t]
    \centering
    \includegraphics[width=\textwidth]{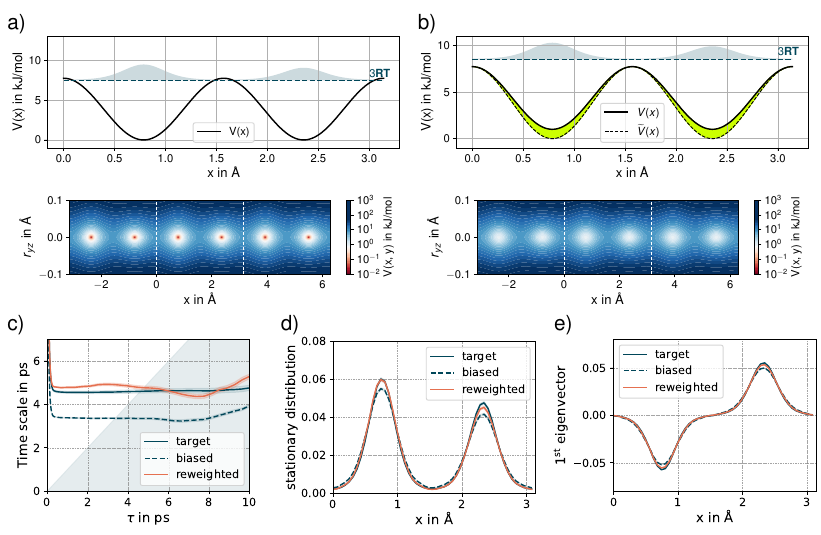}  
    \caption{Markov State Model on a 1D-periodic double basin potential: Results of a reweighting study using CP2K and PLUMED.
    a) 1D-profile of the double well showing the unbiased PES and the Boltzmann distribution as well as the $3RT$ percentile of the kinetic energy (top), 2D-image of the unbiased PES periodically extended into $x$, where white dashed lines mark the simulation cell (bottom);
    b) same as a, but after application of the bias potential (shown in yellow);
    c) implied time scales of the slowest process for the biased and unbiased runs as well as after Girsanov reweighting, with
    d) the corresponding stationary distributions, and e) the first eigenvector.
    }
    \label{fig:results_MSM_A}
\end{figure}

Fig.~\ref{fig:results_MSM_A} shows the results of the Girsanov reweighting benchmark for an MSM defined on a periodic double basin potential.
The upper panel illustrates the PES in $x$ with and without the applied bias potential (Fig.~\ref{fig:results_MSM_A}a–b), together with the corresponding 2D representations of the PES.
The associated Boltzmann distributions are shown, including the position of the kinetic-energy percentile $3RT$ relative to the global minimum of the PES. 
According to the Maxwell–Boltzmann energy distribution, 88.8\% of the particles have kinetic energies below this value.
Therefore, a barrier substantially higher than $3RT$ can be considered linked to a rare event, as it is more probable that a particle reaching this point in the PES will relax back to the well than cross the barrier. 
In the unbiased system (Fig.~\ref{fig:results_MSM_A}a), the barrier height is low enough to ensure rapid convergence of the reference data. 
Upon applying the bias potential, the barrier is effectively reduced such that $3RT$ lies above the barrier.

The stationary distribution extracted from the MSM is consistent with the calculated Boltzmann distribution for both the unbiased and the biased runs (Fig.~\ref{fig:results_MSM_A}d).  
A stable implied timescale of 4.6~ps
is found for the slowest process in the unbiased run (Fig.~\ref{fig:results_MSM_A}c) and the first eigenvector of the MSM proves that this value belongs to the barrier crossing (Fig.~\ref{fig:results_MSM_A}e).
This process is significantly accelerated through the bias, exhibiting an implied time scale of 3.4~ps (74\%).
The stationary distribution and the first eigenvector, which still represents the barrier crossing, are also broadened in accordance with the theoretical Boltzmann distribution.

Reweighting the MSM from the biased potential to the (unbiased) target potential is successful and reproduces the implied time scale of the unbiased simulation (Fig.~\ref{fig:results_MSM_A}c).
The stationary distribution and the first eigenvector--although only slightly perturbed by the bias potential--are also correctly recovered through reweighting (Fig.~\ref{fig:results_MSM_A}d–e).
The statistical uncertainty of the timescales of the biased run is significantly smaller than that of the target run, because it resembles the more diffusive process, which improves the sampling.
Upon reweighting, however, an uncertainty close to that of the target run is obtained.
This is noteworthy given that the applied bias is comparatively weak, and it is reasonable to expect that stronger biasing forces would introduce even more noise upon reweighting, which therefore requires higher sampling.
However, while this model primarily serves to demonstrate the feasibility of Girsanov reweighting with the new CP2K implementation, it should be noted that it is a simple model:
The path stability of a single particle moving freely on the PES is much lower than that of a collective variable in a real system.
This follows from its lower mass and thus low inertia with respect to all acting forces, including the biasing force.
Indeed, reducing the mass of the particle not only reduces the time scales of the barrier crossing, but also renders Girsanov reweighting inefficient (see Fig.~\ref{fig:results_MSM_A_LowMass}a in Appendix~\ref{app:results}).

\begin{figure}[h]
    \centering
    \includegraphics[width=\textwidth]{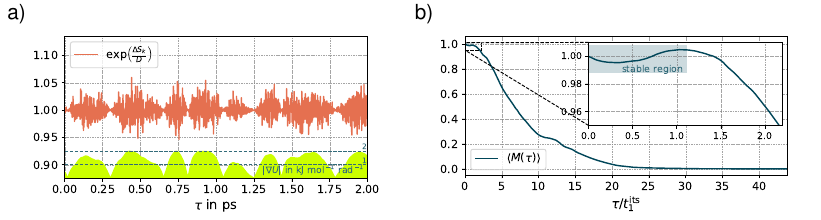}    
    \caption{Assessment of the stability of the reweighting of the MSM shown in Fig.~\ref{fig:results_MSM_A}.
    a) Single-particle example of the time evolution of the increments used for computing the dynamical reweighting.
    b) Average dynamical reweighting factor of all particles as a function of lag time $\tau$ divided by the implied time scale of the unbiased system.}
    \label{fig:M_average}
\end{figure}
Fig.~\ref{fig:M_average} (together with its pendants, Fig.~\ref{fig:results_MSM_A_LowMass}b-c in Appendix~\ref{app:results}) illustrates the effect of path weight stability on the dynamical reweighting factor $M(\tau)$.
For individual trajectories, the increments $\exp\left(\tfrac{\Delta S_k}{D}\right)$ indexed by time step $k$ oscillate about unity (i.e., equal path weights) as shown in Fig.~\ref{fig:M_average}a.
The regions where the biasing force acts (shown in yellow) are readily identified by pronounced oscillations of the increments, in contrast to regions with a flat bias potential, which exhibit only weak oscillations. 
This behaviour follows directly from the definition of the reweighting factor, which depends on the magnitude of $\nabla{U}(\bm{q})$.
The dynamical reweighting factor $M$ is obtained as the cumulative product of increments (Eq.~\ref{eq:logM}) and feeds mainly on regions with strong oscillations. 
$M$ will inevitably decay to zero; the longer a path runs under $V(\bm{q})$, the less likely it is to be found under $\widetilde{V}(\bm{q})$.
However, the extent of the \textit{stable region} preceding this decay defines the effective window for Girsanov reweighting (Fig.~\ref{fig:M_average}b).
When the strongly oscillatory regions occur as well-separated segments, like in the example shown in Fig.~\ref{fig:M_average}a, $M$ retains its path memory, which extends the stable region.
For the present case, the window of stable path weights exceeds the implied time scale of the unbiased system and allows for efficient reweighting.

In contrast, when the inertia of the particle (or collective variable) is low, $M$ is overloaded with too many transitions that require frequent path adjustments at the same time (Fig.~\ref{fig:results_MSM_A_LowMass}b-c in Appendix~\ref{app:results}).
As a result, $M$ decays too early at about 50~\% of the implied time scale of the slowest process, rendering Girsanov reweighting inefficient.

In addition to the effective mass of the particle (or collective variable), several other parameters can enhance the stability of $M$, such as 
a lower temperature, 
a larger friction coefficient $\xi$,
or a weaker biasing force. 
Ultimately, these factors must be balanced against sampling efficiency and the degree to which the thermostat interferes with the underlying dynamics, a detailed investigation of which lies beyond the scope of this study.
It is important to note that the time window cannot be significantly extended by increasing the number of samples, as it is determined by the intrinsic connection of the increments to the path dynamics.

% ----------------------------------------------------------------------------
\subsubsection{Diffusion in a Lennard-Jones Liquid}
\begin{figure}[ht]
    \centering
    \includegraphics[width=\textwidth]{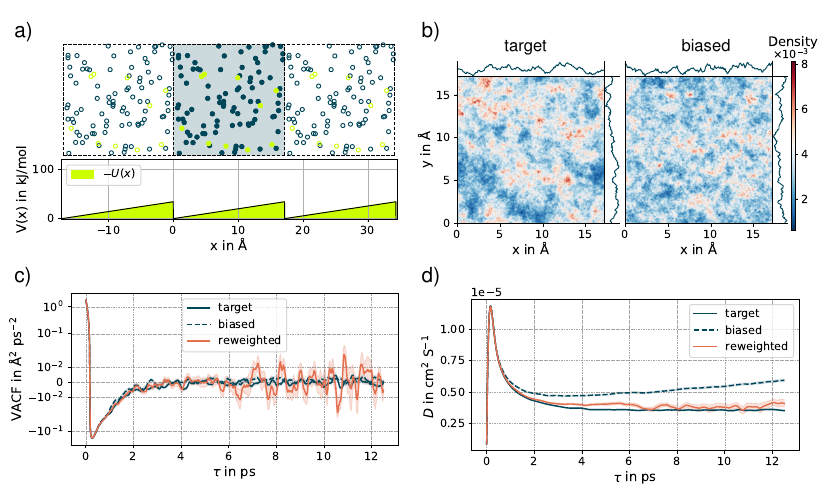}%
    \caption{Diffusion in a Lennard-Jones Liquid: Results of reweighting the diffusion constant after application of an external linear--periodic (saw-tooth) potential.
    a) System setup with the simulation cell (light blue) as well as two periodic images in $x$. Particles highlighted in yellow are biased with the potential shown in the lower panel;
    b) 2D image of the averaged spatial probability density of the particles for the unbiased (left) and the biased (right) simulations (summed over $z$);
    c) velocity autocorrelation function (VACF) of biased and unbiased particles, as well as the result of the reweighting (note the logarithmic scale in $y$);
    d) diffusion coefficent as cumulative integral over the VACF.}
    \label{fig:results_Diffusion}
\end{figure}
Fig.~\ref{fig:results_Diffusion} summarizes the results obtained from reweighting the diffusion constant of a LJ liquid (argon). 
The simulation setup is illustrated in the upper panel: for every run, a linear–periodic bias potential in the $x$-direction is applied to 10 out of 108 particles (highlighted in yellow), resulting in a constant biasing force (Fig.~\ref{fig:results_Diffusion}a). 
Because periodic boundary conditions are used, particles exiting the simulation box on one side reenter on the opposite side. 
Consequently, when a linear bias potential is imposed, no preferred spatial region in $x$ emerges for locating the biased particles. 
This is confirmed numerically by comparing the spatial distributions of particles in biased and unbiased simulations that show a largely uniform distribution (Fig.~\ref{fig:results_Diffusion}b).
As a consequence, the static reweighting factor $g(\bm{x}_0)$ can (or must) be set to one.

The VACF shown in Fig.~\ref{fig:results_Diffusion}c is very similar for both the unbiased and biased runs, as expected for a weak biasing force. 
Still, after the first minimum, the biased VACF decays to zero more quickly. 
The reweighted VACF clearly approaches the target run, but becomes numerically unstable after approximately 5~ps.
A more pronounced difference between the simulations emerges upon integration of the VACF, which yields the diffusion coefficient (Fig.~\ref{fig:results_Diffusion}d). 
While the unbiased run converges to a constant value of 0.35~cm$^2$~s$^{-2}$, the biased run diverges linearly toward increasingly large values.
This behaviour arises from the constant application of the biasing force, which introduces a correlated velocity drift and leads to a linear increase in $D$.
Even though, the biased system thus resembles a non-equilibrium situation with a constant flux of particles in $x$, the velocity drift can be reset and dynamical reweighting is possible recovering the target value.
After 6~ps, the numerical uncertainty increases significantly.
This is the first time dynamical reweighting involving particle velocities has been carried out. 
For underdamped Langevin dynamics, phase space paths are defined by both positions and velocities, which explains why this result is expected.
A non-equilibrium situation could be avoided by using a more sophisticated bias potential. For example, one could define biased and unbiased regions.

\begin{figure}[h]
    \centering
    \includegraphics[width=\textwidth]{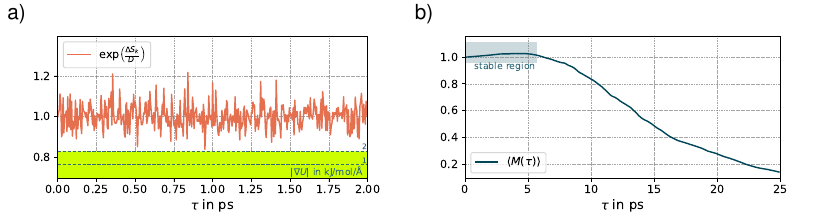}            
    \caption{Assessment of the stability of the reweighting of the diffusion constant shown in Fig.~\ref{fig:results_Diffusion}.
    a) Single-particle example of the time evolution of the increments used for computing the dynamical reweighting.
    b) Average dynamical reweighting factor of all particles as a function of lag time $\tau$.}
    \label{fig:M_average_diffusion}
\end{figure}
As for the MSM, the stability of $M$ can be examined by analysing the oscillations of the increments in$\exp\left(\tfrac{\Delta S_k}{D}\right)$ and their average cumulative product (Fig.~\ref{fig:M_average_diffusion}).
Because the biasing force is constant, the increments fluctuate with similar amplitude around unity.
$M(\tau)$ still decays rapidly after approximately 6~ps, which corresponds to the time window where the reweighted VACF remained numerically stable.
In this case as well, stronger intrinsic friction, effectively increasing the particle’s mass, could stabilise the diffusion constant in the biased run, thus extending the time window over which Girsanov reweighting remains reliable.
Yet a study of this is beyond the scope of this feasibility study.

% ----------------------------------------------------------------------------
\section{\label{sec:Conclusion}Conclusion}
% ----------------------------------------------------------------------------

We have presented the successful implementation of Girsanov reweighting for molecular dynamics simulations within the general-purpose MD/DFT code CP2K.
We demonstrated that dynamical quantities could be reweighted successfully.
This has been demonstrated for Markov state models, as well as for transport properties by reweighting of the diffusion coefficient.
Upon benchmarking our implementation, we also analysed the stability of the path weights that define the time window for efficient Girsanov reweighting.
Notably, we found that these measures converge more quickly than the actual observables, allowing for an early evaluation of the quality of the reweighting bias.

The CP2K implementation can be used immediately for a range of applications where biased sampling plays a central role, such as catalytic cycle optimisation or reaction pathway exploration.\cite{yang_molecular_2025,ketkaew_closer_2022,iannuzzi_cp2k_2025, vidossich_role_2021,weis_exploring_2020,bueno_ethanol_2023}
Another important application area is the reweighting of low-frequency vibrational modes in the terahertz (THz) range, which are also referred to as large-amplitude motions (LAMs). 
Although AIMD is well suited to capturing fast vibrational motions, slow conformational dynamics remain severely undersampled due to its high computational cost.\cite{B006,R053,R013}
On the other hand, spectroscopies such as vibrational circular dichroism (VCD) are highly sensitive to the underlying conformational populations, making comprehensive sampling essential.\cite{P009,P019,P014}
The application of a bias potential along LAMs during conformational sampling alters the vibrational density of states (vDOS) in that region, and may even affect vibrational modes of a higher frequency.\cite{G160,P019}
This exerts an indirect influence on the IR and VCD spectra and requires dynamical reweighting.

However, whether spectroscopic and transport properties can be reliably reweighted depends on the sensitivity of the corresponding dynamical observables to the thermostatting scheme.
Using the Green-Kubo formalism, Girsanov reweighting can be employed as a variance reduction method to calculate the transport properties of stochastic dynamics \cite{gastaldello2025dynamical}. 
However, when evaluating physical transport properties such as diffusion coefficients, vibrational densities of state, and viscosities, the relevant dynamical observables are modified not only by the bias potential, but also by the stochastic MD integrator itself.\cite{G213, G194, G088}
Thus, Girsanov reweighting cannot be used as it is to obtain transport coefficients.
In Ref.~\citenum{heinz_how_2025}, we examined how the thermostat coupling time compares with the correlation function decay timescale and the stochastic environmental forces acting on the relevant collective variable. 
We found that if the thermostat couples more weakly than the environmental fluctuations, its distortion of the correlation function is negligible.
Thus, in these cases reweighting of physical transport properties should be possible.
Another way of minimising dynamical distortions would be to improve the design of thermostatted regions.
Currently, CP2K only offers multiple defined regions with individually global thermostats (e.g.~MOLECULE), but not with individually massive thermostats.
Applying a massive thermostat to the biased subspace and a global thermostat to the remainder of the system could provide a best-of-both-worlds solution and improve the quality of reweighted dynamical properties using Girsanov reweighting.

% TO BE IMPLEMENTED (LATER)
% 	-- combination of biases:
% 		--- allow for repeated &BIAS sections
% 		--- esp. for EXTERNAL+RESTRAINT make sure that there is no double counting of U! (both use force_env/additional_potential)
% 		--- also requires update of LOG output
% 		--- we need ONE delta S_k per bias (so if we have 100 biases, we need 100 factors)
% 			----> maybe we do not do it, because in fact one can construct everything from what is there
% 				(eta output + U from different sources like PLUMED)
% 	-- MODE PERTURBATION/LINRES 
% 	-- MIXING (force_env_methods.F:mixed_energy_forces)
% 	-- other ensembles (LANGEVIN, NPT etc.)

% 	-- thermostat for atomlist only (rest global) == defined region

% 	-- ATOM_LIST, could be along list (use dots like in QM/MM) ...
%           IF (cons_info%g33_intermolecular(ig)) THEN
%            CALL section_vals_val_unset(g3x3_section, "MOLECULE", i_rep_section=ig)
%            CALL section_vals_val_unset(g3x3_section, "MOLNAME", i_rep_section=ig)

% ----------------------------------------------------------------------------
\section*{References}
% ----------------------------------------------------------------------------

% \nocite{*}
\bibliographystyle{vancouver}
\bibliography{references}  %aipsamp}% Produces the bibliography via BibTeX.

\begin{acknowledgments}
This research has been funded 
by Deutsche Forschungsgemeinschaft (DFG) through grant SFB 1114 "Scaling Cascades in Complex Systems" (project number 235221301, project B05: "Origin of scaling cascades in protein dynamics"), and 
by the Volkswagen Foundation through a Momentum grant.
Artificial intelligence tools (ChatGPT and DeepL for text optimisation and ChatGPT for figure generation using matplotlib) were used to help prepare this manuscript. All AI-generated content was reviewed, edited, and approved by the authors, who assume full responsibility for the final work.
\end{acknowledgments}

\section*{Data Availability}

Data for this article, including input files and analysis scripts are available via Zenodo at \href{https://doi.org/10.5281/zenodo.17964423}{https://doi.org/10.5281/zenodo.17964423}.
The source codes of the development version of CP2K is publicly available via the author's GitHub page \href{https://github.com/sjaehnigen/cp2k}{https://github.com/sjaehnigen/cp2k} (branch: \texttt{girsanov\_reweighting}). 

\section*{Conflict of Interests}
 There are no conflicts to declare. 
 
\section*{Appendix}
\appendix

\section{Girsanov Reweighting in the O'V'RV'O' scheme}
\label{app:OVRVO}

See Ref.~\citenum{X001} for further details.
The \textit{total} update operator is obtained by applying update steps sequentially.
\begin{align*}
&\mathcal{O'V'RV'O'}\binom{q_{i,k}}{\dot{q}_{i,k}} \\
& =\mathcal{O'V'RV'}\binom{q_{i,k}}{d' \dot{q}_{i,k}+f' \eta_{i,k}^{(1)}} \\
& =\mathcal{O'V'R}\binom{q_{i,k}}{d' \dot{q}_{i,k}+f' \eta_{i,k}^{(1)}+b_i'\left(\bm{q}_{k}\right)} \\
& =\mathcal{O'V'}\binom{q_{i,k}+a d' \dot{q}_{i,k}+a f' \eta_{i,k}^{(1)}+a b_i'\left(\bm{q}_{k}\right)}{d' \dot{q}_{i,k}+f' \eta_{i,k}^{(1)}+b_i'\left(\bm{q}_{k}\right)} \\
& =\mathcal{O'}\binom{q_{i,k}+a d' \dot{q}_{i,k}+a f' \eta_{i,k}^{(1)}+a b_i'\left(\bm{q}_{k}\right)}{d' \dot{q}_{i,k}+f' \eta_{i,k}^{(1)}+b_i'\left(\bm{q}_{k}\right)+b_i'\left(\bm{q}_k+a d' \dot{\bm{q}}_k+a f' \bm{\eta}_k^{(1)}+a b_i'\left(\bm{q}_{k}\right)\right)} \\
& =\binom{q_{i,k}+a d' \dot{q}_{i,k}+a f' \eta_{i,k}^{(1)}+a b_i'\left(\bm{q}_{k}\right)}{d' d' \dot{q}_{i,k}+d' f' \eta_{i,k}^{(1)}+d' b_i'\left(\bm{q}_{k}\right)+d' b_i'\left(\bm{q}_k+a d' \dot{\bm{q}}_k+a f' \bm{\eta}_k^{(1)}+a b_i'\left(\bm{q}_{k}\right)\right) + f' \eta_{i,k}^{(    2)}},
\end{align*}
which yields the \textit{update operator} $\mathcal{U}_{\mathcal{OVRVO}}: \mathbb{R}^2\times\Gamma \to\Gamma$
\begin{align}
\binom{q_{i,k+1}}{\dot{q}_{i,k+1}} & =\mathcal{U}_{\mathcal{OVRVO}}\left(\eta_{i,k}^{(1)}, \eta_{i,k}^{(2)} ; \bm{x}_{k}, V\right) \cr
& =\binom{\bar{q}_{i,k+1}}{\bar{\dot{q}}_{i,k+1}}
+ \binom{0}{d' b_i'\left(\bar{\bm{q}}_{k+1}+a f' \bm{\eta}_{k}^{(1)}\right)}
+ \binom{a f'}{d' f'} \eta_{i,k}^{(1)}
+ \binom{0}{f'} \eta_{i,k}^{(2)}.
\label{eq:update_function}
\end{align}
The deterministic part of the update moves the system to the support point $(\bar{q}_{i,k+1}, \bar{\dot{q}}_{i,k+1})$
with $\bar{q}_{i,k+1}=q_{i,k}+a d' \dot{q}_{i,k}+a b_i'\left(\bm{q}_{k}\right)$ 
and $\bar{\dot{q}}_{i,k+1}=d' d' \dot{q}_{i,k}+d' b_i'\left(\bm{q}_{k}\right)$. 
% Thus,
% $$
% \begin{array}{rlrl}
% \mathcal{U}_{\mathcal{OVRVO}}: & \mathbb{R}^{2} & \rightarrow \Gamma \\
% \mathcal{U}_{\mathcal{OVRVO}}:\left(\eta_{i,k}^{(1)}, \eta_{i,k}^{(2)}\right) & \mapsto x_{i,k+1}
% \end{array}
% $$
%
%
%
$\Delta \eta_{i,k}^{(1)}$ and $\Delta \eta_{i,k}^{(2)}$ can be derived from the condition
\begin{equation}
\binom{0}{0}=\mathcal{U}_{\mathcal{OVRVO}}\left(\eta_{i,k}^{(1)}, \eta_{i,k}^{(2)} ; \bm{x}_{k}, V\right)-\mathcal{U}_{\mathcal{OVRVO}}\left(\widetilde{\eta}_{i,k}^{(1)}, \widetilde{\eta}_{i,k}^{(2)} ; \bm{x}_{k}, \widetilde{V}\right) 
\end{equation}
Inserting Eq.~\ref{eq:update_function} yields
\begin{align*}
\binom{0}{0} 
=&\binom{q_{i,k}+a d' \dot{q}_{i,k}+a b_i'\left(\bm{q}_{k}\right)}{d' d' \dot{q}_{i,k}+d' b_i'\left(\bm{q}_{k}\right)}
+ \binom{0}{d' b_i'\left(\bm{q}_{k+1}\right)}
+ \binom{a f'}{d' f'} \eta_{i,k}^{(1)}+\binom{0}{f'} \eta_{i,k}^{(2)} \\
&-\binom{q_{i,k}+a d' \dot{q}_{i,k}+a \widetilde{b}_i'\left(\bm{q}_{k}\right)}{d' d' \dot{q}_{i,k}+d' \widetilde{b_i'}\left(\bm{q}_{k}\right)}
- \binom{0}{d' \widetilde{b}_i'\left(\widetilde{\bm{q}}_{k+1}\right)}
- \binom{a f'}{d' f'} \widetilde{\eta}_{i,k}^{(1)}
- \binom{0}{f'} \widetilde{\eta}_{i,k}^{(2)} \\
= & \binom{a \frac{\Delta t}{2} \nabla_i U\left(\bm{q}_{k}\right)}{d' \frac{\Delta t}{2} \nabla_i U\left(\bm{q}_{k}\right)}
+ \binom{0}{d' b_i'\left(\bm{q}_{k+1}\right)-d'\widetilde{b_i'}\left(\widetilde{\bm{q}}_{k+1}\right)}
- \binom{a f'}{d' f'} \Delta \eta_{i,k}^{(1)}-\binom{0}{f'}\Delta \eta_{i,k}^{(2)}, 
\end{align*}
which follows from Eqs.~\ref{eq:V_tilde} and \ref{eq:parameters_half}.
The second term evaluates the potential at the updated positions which, a priori, might differ in $V$ and $\widetilde{V}$. Solving the line for the position in the above equation yields
\begin{equation}
\underline{\Delta \eta_{i,k}^{(1)}=\frac{1}{f'} \frac{\Delta t}{2} \nabla_i U\left(\bm{q}_{k}\right) }
\end{equation}
With this $\bm{q}_{k+1}=\widetilde{\bm{q}}_{k+1}$, and thus $d' b_i'\left(\bm{q}_{k+1}\right)-d' \widetilde{b_i'}\left(\widetilde{\bm{q}}_{k+1}\right)=d' \frac{\Delta t}{2} \nabla_i U\left(\bm{q}_{k+1}\right)$. 
Then the line for the momentum yields
\begin{align}
& 0=d' \frac{\Delta t}{2} \nabla_i U\left(\bm{q}_{k}\right)+d' \frac{\Delta t}{2} \nabla_i U\left(\bm{q}_{k+1}\right)-d' f' \cdot \frac{1}{f'} \frac{\Delta t}{2} \nabla_i U\left(\bm{q}_{k}\right)-f' \Delta \eta_{i,k}^{(2)} \cr
& =d' \frac{\Delta t}{2} \nabla_i U\left(\bm{q}_{k+1}\right)-f' \Delta \eta_{i,k}^{(2)} \cr
& \Updownarrow \cr
& \underline{\Delta \eta_{i,k}^{(2)}=\frac{d'}{f'} \frac{\Delta t}{2} \nabla_i U\left(\bm{q}_{k+1}\right) }
\end{align}

\section{Further Results}

\label{app:results}

\begin{figure}[ht]
    \centering
    \includegraphics[width=\textwidth]{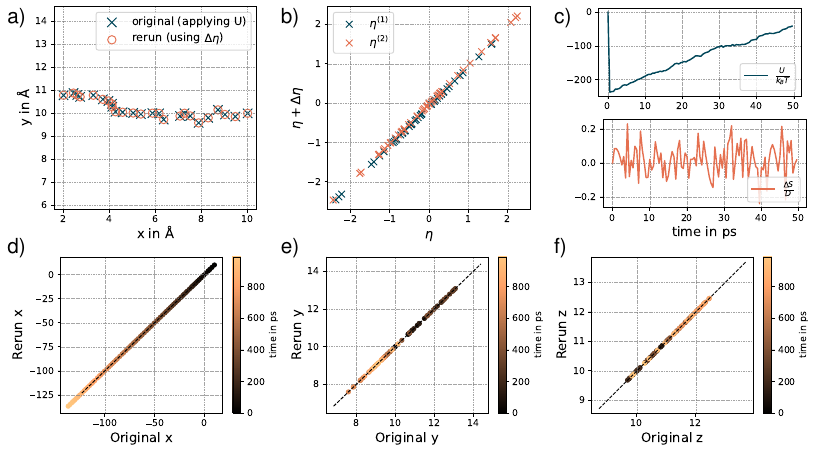}
    \caption{Simple Rerun Benchmarks: Results of the rerun benchmark of a Lennard-Jones particle biased with a linear potential in $x$ using PLUMED.
    a) Superposition of the trajectories generated with CP2K applying the bias $-U(\bm{q})$ and the ones generated with the $\mathcal{O'V'RV'O'}$ script using the $\Delta\eta^{(z)}$ provided by the CP2K run;
    b) Correlation of $\eta^{(z)}$ and $\widetilde{\eta}^{(z)}=\eta^{(z)} + \Delta\eta^{(z)}$;
    c) Static (blue) and dynamic (orange) reweighting factor increments in their logarithmic form;
    d-f) Correlation of the positions in $x$, $y$, $z$, respectively, between the original run and the rerun. The color indicates the time step (black: starting point).}
    \label{fig:results_rerun_PLUMED_l}
\end{figure}

\begin{figure}[ht]
    \centering
    \includegraphics[width=\textwidth]{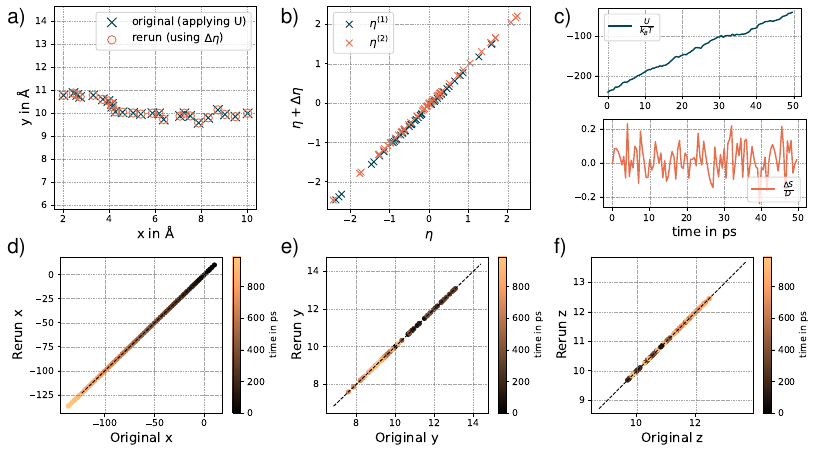}
    \caption{Simple Rerun Benchmarks: Results of the rerun benchmark of a Lennard-Jones particle biased with a linear potential in $x$ using \texttt{EXTERNAL\_POTENTIAL} of CP2K.
    a) Superposition of the trajectories generated with CP2K applying the bias $-U(\bm{q})$ and the ones generated with the $\mathcal{O'V'RV'O'}$ script using the $\Delta\eta^{(z)}$ provided by the CP2K run;
    b) Correlation of $\eta^{(z)}$ and $\widetilde{\eta}^{(z)}=\eta^{(z)} + \Delta\eta^{(z)}$;
    c) Static (blue) and dynamic (orange) reweighting factor increments in their logarithmic form;
    d-f) Correlation of the positions in $x$, $y$, $z$, respectively, between the original run and the rerun. The color indicates the time step (black: starting point).}    
    \label{fig:results_rerun_external}
\end{figure}

\begin{figure}[ht]
    \centering
    \includegraphics[width=\textwidth]{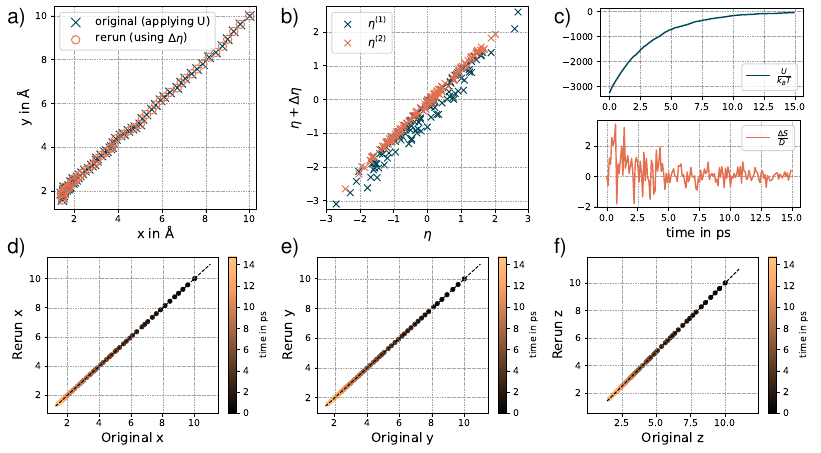}
    \caption{Simple Rerun Benchmarks: Results of the rerun benchmark of a Lennard-Jones particle forced into a potential well at $(x, y) = (1, 1)$ using \texttt{RESTRAINT} of CP2K.
    a) Superposition of the trajectories generated with CP2K applying the bias $-U(\bm{q})$ and the ones generated with the $\mathcal{O'V'RV'O'}$ script using the $\Delta\eta^{(z)}$ provided by the CP2K run;
    b) Correlation of $\eta^{(z)}$ and $\widetilde{\eta}^{(z)}=\eta^{(z)} + \Delta\eta^{(z)}$;
    c) Static (blue) and dynamic (orange) reweighting factor increments in their logarithmic form;
    d-f) Correlation of the positions in $x$, $y$, $z$, respectively, between the original run and the rerun. The color indicates the time step (black: starting point).}    
    \label{fig:results_rerun_restraint}
\end{figure}

\begin{figure}[ht]
    \centering
    \includegraphics[width=\textwidth]{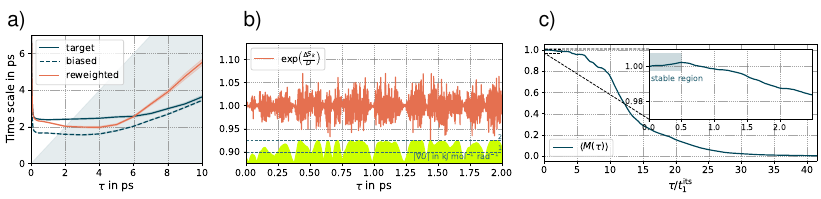}%
    \caption{Markov State Model on a 1D-periodic double basin potential using a lower particle mass (30~$m$-\%) as well as a stronger thermostat (50~$\tau$-\%) (see also Fig.~\ref{fig:results_MSM_A}): Results of a reweighting study using CP2K and PLUMED.
    a) Implied time scales of the slowest process for the biased and unbiased runs as well as after Girsanov reweighting;
    b) single-particle example of the time evolution of the increments used for computing the dynamical reweighting;
    c) average dynamical reweighting factor of all particles as a function of lag time $\tau$.}
    \label{fig:results_MSM_A_LowMass}
\end{figure}

\end{document}